\documentclass[useAMS,usenatbib]{mn2e}

\usepackage{txfonts,graphicx,amssymb,subfigure}
\usepackage[normalem]{ulem}
\usepackage{color}

\title{Faraday Rotation Measure Synthesis for Magnetic Fields of Galaxies}

\author[P.~Frick et al.]
       {P.~Frick$^1$, D.~Sokoloff$\,^{2}$,  R.~Stepanov$^1$,
        and R.~Beck$^3$\\
$^1$ Institute of Continuous Media Mechanics,
Korolyov str.~1, 614013 Perm, Russia \\
$^2$ Department of Physics, Moscow University, 119899, Moscow, Russia \\
$^3$ Max-Planck-Institut f\"ur Radioastronomie, Auf dem H\"ugel 69,
  53121 Bonn, Germany}

\date{Accepted 2010 .... Received 2010 ....; in original form 2010}

\pagerange{\pageref{firstpage}--\pageref{lastpage}}
\pubyear{2009}
\begin{document}
\maketitle

\label{firstpage}

%
\begin{abstract}

RM Synthesis was recently developed as a new tool for the
interpretation of polarized emission data in order to separate the
contributions of different sources lying on the same line of sight.
Until now the method was mainly applied to discrete sources in
Faraday space (Faraday screens). Here we consider how to apply RM
Synthesis to reconstruct the Faraday dispersion function,
aiming at the further extraction of
information concerning the magnetic fields of extended sources, e.g.
galaxies. The main attention is given to two related novelties in
the method, i.e. the symmetry argument in Faraday space and the
wavelet technique.

We give a relation between our method and the previous applications
of RM Synthesis to point-like sources. We demonstrate that the
traditional RM Synthesis for a point-like source indirectly  implies
a symmetry argument and, in this sense, can be considered as a
particular case of the method presented here. Investigating the
applications of RM Synthesis to polarization details associated with
small-scale magnetic fields, we isolate an option which was not
covered by the ideas of the Burn theory, i.e. using quantities
averaged over small-scale fluctuations of magnetic field and
electron density. We describe the contribution of small-scale
fields in terms of Faraday dispersion and beam depolarization. We
 consider the complex polarization for RM Synthesis without any
averaging over small-scale fluctuations of magnetic field and
electron density and demonstrate that it allows us to
 isolate the contribution from small-scale field.

A general conclusion concerning the applicability of RM Synthesis to
the interpretation of the radio polarization data for extended
sources, like galaxies, is that quite severe requirements
 (in particular to the wavelength range covered by
observations) are needed to recognize at least the principal structure of the Faraday dispersion function. If
the wavelength range of observations is not adequate we
describe which features of this function can be
reconstructed.

\end{abstract}


\section{Introduction}
\label{intro}

RM Synthesis is a new tool for the interpretation of polarized
emission data in order to get information on the emitting media
\citep{2005A&A...441.1217B}. Until now the method was mainly applied
to discrete sources in Faraday space (``Faraday screens'')
\citep{2009IAUS..259..591H}. The aim of this paper is to discuss applications
to sources which are extended in Faraday space.

RM Synthesis is based on the fact that
the complex polarized intensity $P$ can be calculated as a
Fourier transform
\begin{equation}
\label{p_to_f} P(\lambda^2) =  \int_{-\infty}^{\infty} F(\phi)
{\rm e}^{2{\rm i}\phi \lambda^2}  \mathrm{d} \phi
\end{equation}
of the Faraday dispersion function $F$ in Faraday space with the
coordinate $\phi$ which is the Faraday depth
\citep{Burn1966MNRAS.133...67B}. Performing the inverse Fourier
transform of $P$ one obtains $F$ which is the polarized intensity
emerging from a region with Faraday depth $\phi$. Faraday depth is
defined as
\begin{equation}
\phi(x) = 0.81\int_{0}^{x} B_\parallel
(x') n_{\rm e}(x') \mathrm{d}x', \label{fardep}
\end{equation}
where $B_\parallel$ is the line-of-sight magnetic field component
measured in $\mu$G, $n_{\rm e}$ is the thermal electron density measured
in cm$^{-3}$ and the integral is taken from the observer to the
current point $x$ along the line of sight over the region which
contains both magnetic fields and thermal electrons, where $x'$ is
measured in parsecs. A convention is that $\phi$ is positive
when $\mathbf{B}$ is pointing towards the Earth.

In the context of RM Synthesis, one has to distinguish between
Faraday depth and Faraday rotation measure $RM$ which is defined as
\begin{equation}
RM (\lambda) ={{\mathrm{d}  \Psi} \over {\mathrm{d} (\lambda^2)}},
\label{RM}
\end{equation}
where $\Psi$ is the polarization angle ($P = |P|{\rm e}^ {2{\rm  i} \Psi}$).

The inversion of formula (\ref{p_to_f}) shows that the Faraday
dispersion function $F$ is the Fourier transform of the complex
polarized intensity:
\begin{equation}
\label{f_to_p1} F(\phi) = {{1} \over{\pi}} \hat P(k), \label{Burn}
\end{equation}
where $k=2\phi$, and the Fourier transform is defined as
\begin{equation}
\label{four} \hat {f}\left( k \right) = \int_{-\infty}^{\infty}
{f\left( y \right){\rm e}^{ -{\rm i} k y} \mathrm{d}y}.
\end{equation}
The practical limitation of the use of Eq.~(\ref{f_to_p1}) comes
from the fact that $P$ can be measured only for $\lambda^2>0$ and in
 only in a finite spectral band \citep{2009IAUS..259..669B}.

The lower bound $\lambda_{\rm min}$ restricts the possibility to
recognize objects which are extended in Faraday space, the upper
bound $\lambda_{\rm max}$ suppresses the visibility of small-scale
structures of the object in Faraday space, and the lack of negative
$\lambda^2$ impedes the correct reconstruction of the intrinsic
polarization angle. Poor sampling inside the
observations window prevent the reconstruction of objects  at large
Faraday depths. All these problems and limitations can also be
illustrated using the wavelet representation, which provides a kind
of local Fourier transform, isolating a given structure on both sides of the
Fourier transform.

The topic of this paper is a further development of the wavelet
approach as suggested by Frick et al. (2010) which presented
at least a partial resolution of these difficulties.
The paper is organized as follows. In Sects.~\ref{wavelet},\ref{rmsf} we describe
the wavelet-based RM Synthesis and the
effect of extending $P$ to negative $\lambda^2$, and in Sect~\ref{clean} we
compare RM-CLEAN \citep{2009IAUS..259..591H} to a wavelet-based RM
deconvolution.  We subsequently describe a
few simplistic, but reasonable models of magnetic fields and
electron densities in galaxy discs. Large-scale features are
described in Sect.~\ref{galactic}, small-scale features in Sect.~\ref{turb}.
The effect of frequency coverage on reconstruction
fidelity is discussed in Sect.~\ref{range}. We subject
these models to traditional RM Synthesis and to our wavelet approach in
Sect.~\ref{SE} and summarize our results in Sect.~\ref{DC}.

\section{Wavelet-based RM Synthesis }
\label{wavelet}

The wavelet transform of the Faraday dispersion function $F(\phi)$
can be written in the form
\begin{equation}
\label{wF_d}
w_F(a,b) = {{1}\over {|a|}} \int\limits_{ - \infty }^\infty
{F(\phi)\psi ^\ast \left( {\frac{\phi - b}{a}} \right)\mathrm{d}\phi} ,
\end{equation}
where $\psi(\phi)$ is the analyzing wavelet, $a$ defines the scale
and $b$ defines the position of the wavelet center.
Then the coefficient $w_F$ gives the contribution of the
corresponding structure of scale $b$ at position $a$ to the function $F$.

The contribution of the complex polarized intensity to the wavelet decomposition of
the Faraday dispersion function  can be divided
in two parts $w_F(a,b) =w_-(a,b)+w_+(a,b)$, which are defined by $P(\lambda^2)$ in negative and positive domain $\lambda^2$, correspondingly. From  Eqs.~(\ref{Burn}) and (\ref{wF_d}) one gets
\begin{equation}
w_+(a,b)={{1}\over {\pi}}
\int\limits_0^\infty {P(\lambda^2){\rm e}^{-2{\rm i}b \lambda^2} \hat{\psi}^\ast
\left( -2a \lambda^2 \right)\mathrm{d}\lambda^2}.
\end{equation}
The coefficients $w_+(a,b)$ are calculated from the known $P(\lambda^2)$ for
$\lambda^2>0$. \cite{2010MNRAS.401L..24F} suggested to
 recognize the dominating structures in the map
$|w_+(a,b)|$. The coordinate $b$ of the corresponding maximum gives
the position $\phi_0$ of the structure in Faraday depth. Then, the coefficients $w_-(a,b)$ can be
reconstructed, following the symmetry arguments (see
Eq.~(\ref{cont}) below)
\begin{equation}
\label{w-w+}
w_-(a,b)=w_+\left(a,2\phi_0(a,b)-b\right),
\end{equation}
where the function $\phi_0(a,b)$ takes on a value of $b$ which
corresponds to the maximum of the dominant structure near point $(a,b)$.
 Distribution of $|w_+(a,b)|$ is usually complex and applying of Eq.~(\ref{w-w+}) can be nontrivial procedure. We suggest it in Secs.~\ref{range} and \ref{SE} how to proceed in some particular cases.

The function $F$ can be restored using the inverse transform
\begin{equation}
\label{wF_inv}
F(\phi) = \frac{1}{C_\psi }\int\limits_{-\infty}^\infty {\int\limits_{ -
\infty }^\infty {\psi \left( {\frac{\phi - b}{a}} \right)w_F\left(
{a,b} \right)\frac{\mathrm{d} a \, \mathrm{d} b}{a^2}} }.
\end{equation}
 We use below the so-called Mexican hat $\psi (\phi) = (1-\phi^2) \exp
(-\phi^2/2)$ as the analyzing wavelet. The wavelet is real, however,
the function $P$ is complex, so the wavelet coefficients $w_F$
are complex as well. For the chosen wavelet $w_F(-a,b)=w_F(a,b)$ and
$C_\psi = 1$.

\section{Rotation Measure Spread Function }
\label{rmsf}

The Faraday dispersion function $\tilde F(\phi)$ obtained by RM Synthesis can be related to the
true Faraday dispersion function $F(\phi)$ by $R(\phi)$, the ``RM
Spread Function'' (RMSF)  is
\begin{equation}
\tilde F(\phi) = F(\phi)\ast R(\phi),
\label{rmsf0}
\end{equation}
where $\ast$ denotes the convolution. The RMSF is defined as
\begin{equation}
R(\phi) = K \int_{-\infty}^{\infty} W(\lambda^2){\rm e}^{-2{\rm i}\phi \lambda^2}\mathrm{d} \lambda^2 ,
\label{rmsf1}
\end{equation}
where $W(\lambda^2)$ is the shape of the observable window in
$\lambda^2$ space (the window function, which can include weights
due to different sensitivities (gains) in the
observation channels, see \cite{2009IAUS..259..591H}) and $K$ is a
normalization constant.
\begin{figure}
\begin{picture}(240,330)(0,0)
\put(0,220){\includegraphics[width=0.4\textwidth]{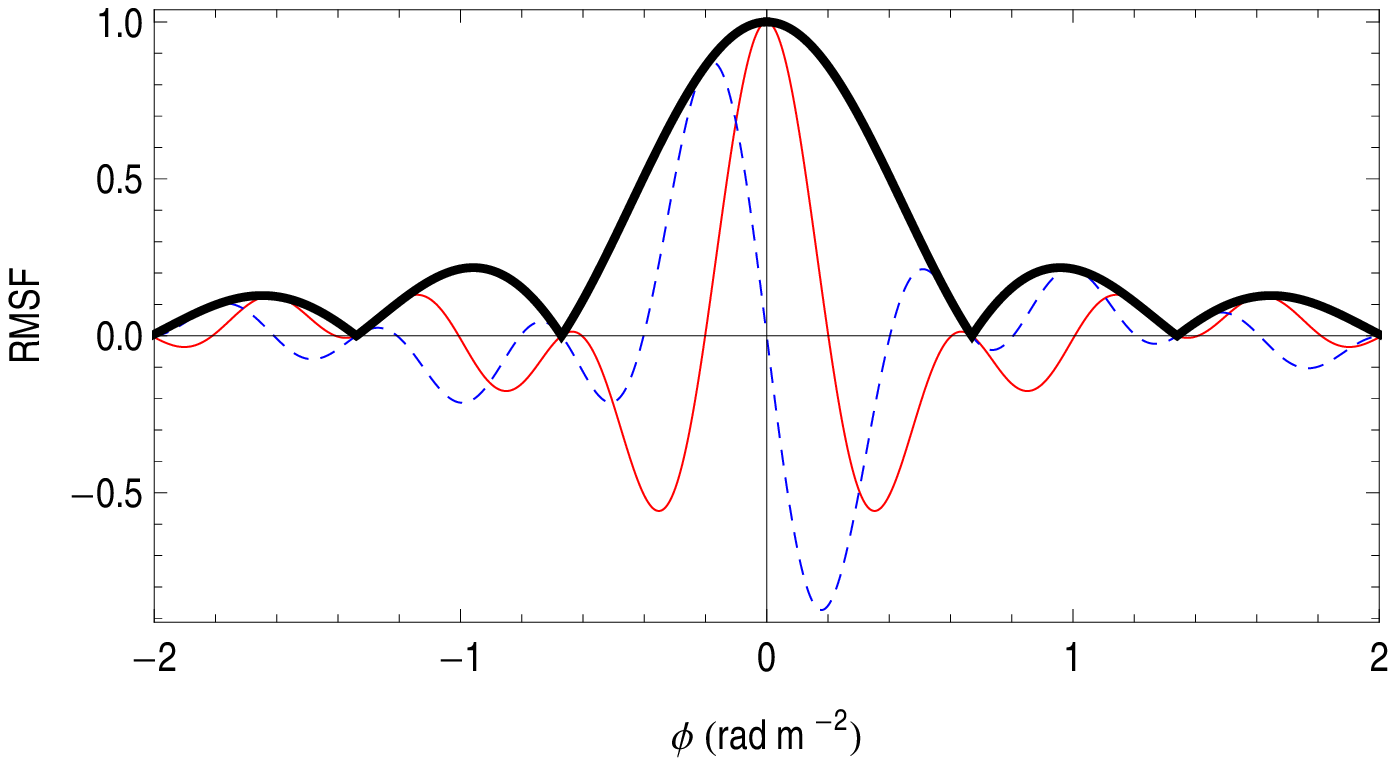}}
\put(0,110){\includegraphics[width=0.4\textwidth]{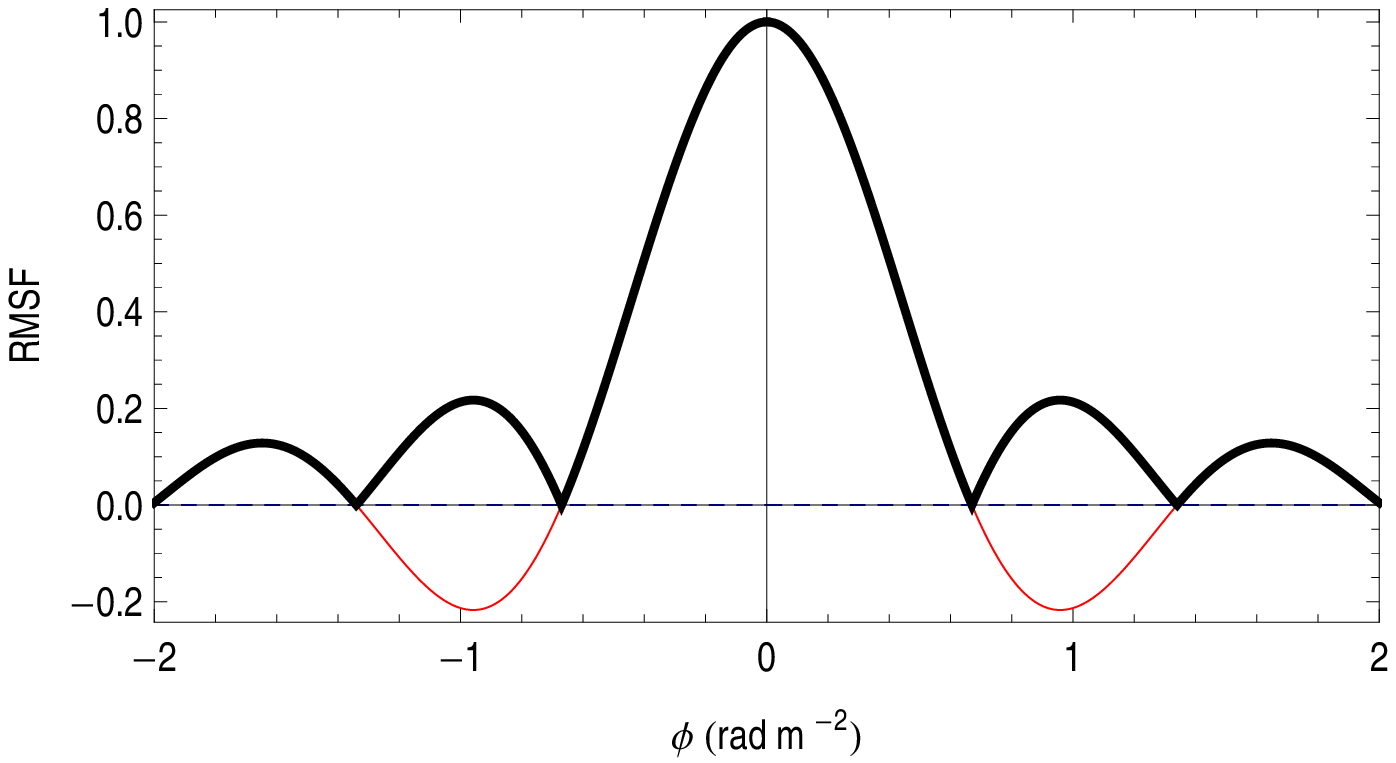}}
\put(0,00){\includegraphics[width=0.4\textwidth]{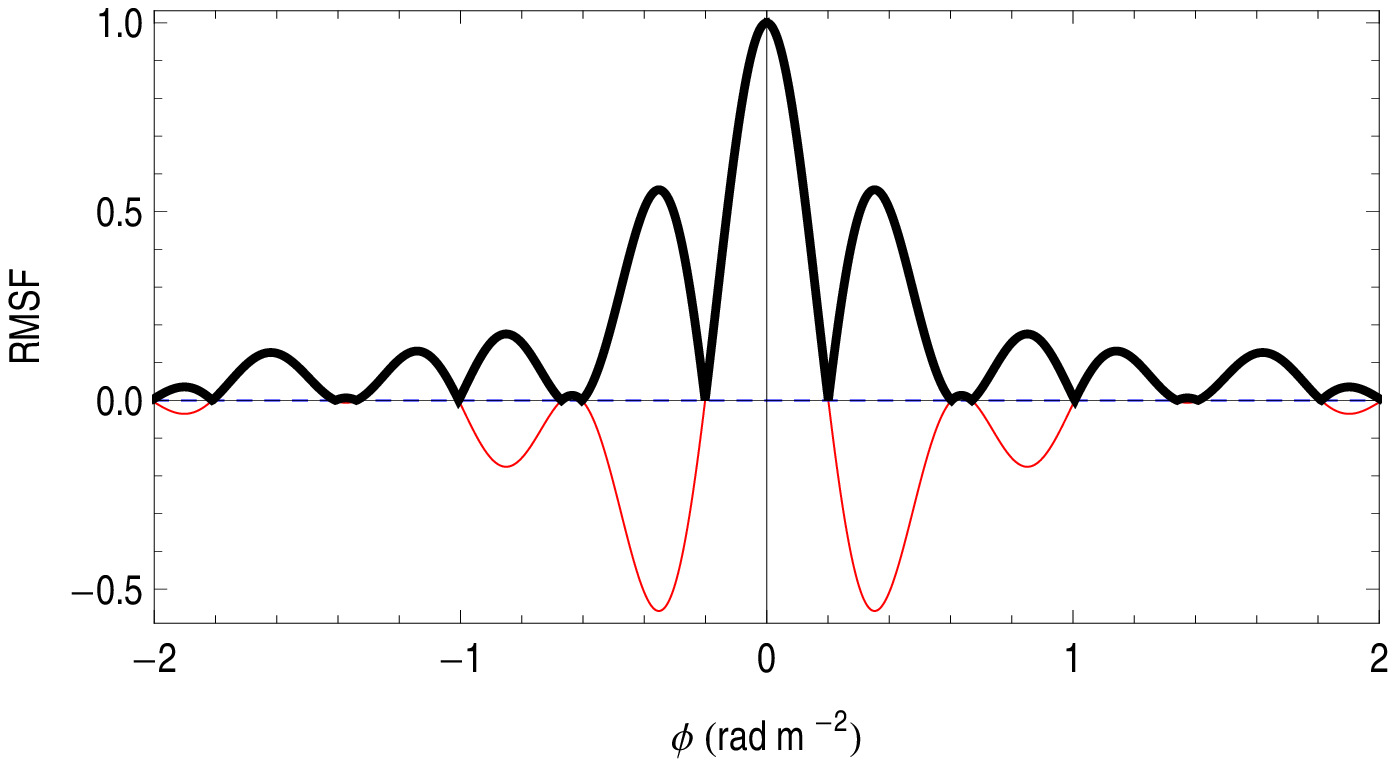}}
\put(210,320){(a)}
\put(210,210){(b)}
\put(210,100){(c)}
\end{picture}
\caption {RM spread functions for a single observation window with
$1.25<\lambda<2.5$\,m. Panel (a): standard $R$; panel (b) $R_{BB}$;
panel (c): $R_W$. Real part (thin solid red), imaginary part (dashed
blue), modulus (thick black).  } \label{fig-rmsf}
\end{figure}

For a single spectral window ($W\equiv1$  for $\lambda_{\rm min} <
\lambda < \lambda_{\rm max}$ and $W\equiv0$ elsewhere) the RMSF has
the simple form
\begin{equation}
R(\phi) = K {\rm e}^{-2{\rm i}\phi \lambda^2_0} \, \frac{\sin{(\phi
\Delta\lambda^2)}}{\phi}, \label{rmsf2}
\end{equation}
where $\lambda_0^2 = (\lambda_{\rm \min}^2 + \lambda_{\rm max}^2)/2$
and $\Delta \lambda^2 = (\lambda_{\rm max}^2 - \lambda_{\rm
min}^2)/2$. $R(\phi)$ is shown in Fig.~\ref{fig-rmsf}a for
$\lambda^2_0=3.91$\, m$^2$ and $\Delta\lambda^2=2.35$\, m$^2$
(corresponding to the LOFAR wavelength range in the
highband). \footnote{For the sake of definiteness, we consider
here and below as a typical example the observational range in the
LOFAR high band ($1.25 < \lambda < 2.5$\, m, see R\"ottgering 2003)
and use another observational range if required to illustrate
particular properties of RM Synthesis.}
The RMSF  usually is a complex-valued function which means that the
obtained Faraday function differs from the true one not only
in amplitude, but also in phase. (Due the lack of data in the domain
of negative $\lambda^2$ RM Synthesis rotates the phase of the
function under reconstruction.) To avoid this,
\cite{2005A&A...441.1217B} proposed to multiply Eq.~(\ref{rmsf2}) by
the factor ${\rm e}^{2{\rm i}\phi \lambda^2_0}$ which means that
\begin{equation}
R_{BB}(\phi) = K \int_{-\infty}^{\infty} W(\lambda^2){\rm e}^{-2{\rm i}\phi
(\lambda^2-\lambda^2_0)}\mathrm{d} \lambda^2  = K\frac{\sin(\phi
\Delta\lambda^2)}{\phi} \label{rmsfBB}\end{equation} and
\begin{equation}
\tilde F(\phi) = {\rm e}^{-2{\rm i}\phi \lambda^2_0}F(\phi)\ast R_{BB}(\phi).
\label{rmsf01}
\end{equation}
Thus, the modified RM spread function $R_{BB}(\phi)$ has a vanishing
imaginary part and a real part, which reproduces the shape of the
envelope of the function $R(\phi)$ (Fig.~\ref{fig-rmsf}b).

Another way to improve the RM spread function follows from
\cite{2010MNRAS.401L..24F}. Their algorithm uses the symmetry
argument (the results of Sect.~5 give strong support for this
argument - realistic objects mainly look like even objects in
Faraday depth space). It means that if the object is centered at
Faraday depth $\phi_0$, then $F(2\phi_0-\phi)=F(\phi)$. In
$\lambda^2$ space this gives
\begin{equation}
P(-\lambda^2) = \exp\left(-4{\rm  i} \phi_0 \lambda^2\right) P(\lambda^2),
\label{cont}
\end{equation}
which means in terms of the spread function for a single window
\begin{equation}
R_{W}(\phi) =  2K\frac{\sin{(\phi \Delta\lambda^2)}}{\phi} \cos{(2\phi \lambda^2_0)}
\label{rmsfW}\end{equation}
and
\begin{equation}
\tilde F(\phi) = {\rm e}^{-4{\rm i}\phi_0\lambda^2_0}F(\phi)\ast R_{W}(\phi).
\label{rmsf00}
\end{equation}
The RM spread function $R_W(\phi)$ is also shown in
Fig.~\ref{fig-rmsf}c. Similar to $R_{BB}$ this function has no
imaginary part, but the real part remains exactly the same as in
Eq.~(\ref{rmsf2}), compare thin solid lines in Fig.\ref{fig-rmsf}a
and Fig.\ref{fig-rmsf}c.

Note that the symmetry assumption  used here works
straightforward only if there is only one source along the light of
sight. If there are several sources the assumption should be applied
using the wavelet technique.

\section{Cleaning and wavelets}
\label{clean}

 We discussed above the reconstruction of objects symmetric in Faraday
space around $\phi=\phi_0$. The simplest
objects for recognition are point-like sources, which can be
described in Faraday space by delta functions
\begin{equation}
F(\phi)=F_0 {\rm e}^{2{\rm i}\chi_0}\delta(\phi-\phi_0).
\label{point_source}
\end{equation}
The corresponding polarized intensity is given by
\begin{equation}
P(\lambda^2)=F_0 {\rm e}^{2{\rm i}\chi_0}{\rm e}^{2{\rm i}\phi_0\lambda^2}.
\label{point_P}
\end{equation}

Note that the relation (\ref{point_P}) determines
$P(\lambda^2)$ for $\lambda^2<0$ and it follows the symmetry
property (\ref{cont}) which is the starting point for our approach.
It means that the traditional RM Synthesis for a point-like source
indirectly  implies this symmetry argument and
can be considered as a particular case of the method presented in this
paper.

To recognize  several point-like sources located on the same line of sight
one can apply a deconvolution procedure using
the Faraday dispersion function, called ``RM-CLEAN''
\citep{2009IAUS..259..591H}. This procedure iteratively finds
the maximum in the reconstructed Faraday dispersion function and
then iteratively subtracts the scaled versions of the RMSF until the
noise level is reached, after which a smoothed representation of the
``CLEAN model'' is used as the approximate true Faraday dispersion
function.

After removing the brightest structure (including the corresponding
sidelobes) the procedure finds the maximum in the remaining data and
restarts the iteration. \cite{2009A&A...503..409H} showed how well
RM-CLEAN deconvolved RM cubes from the WSRT SINGS survey.

\begin{figure}
\begin{picture}(240,330)(0,0)
\put(0,220){\includegraphics[width=0.4\textwidth]{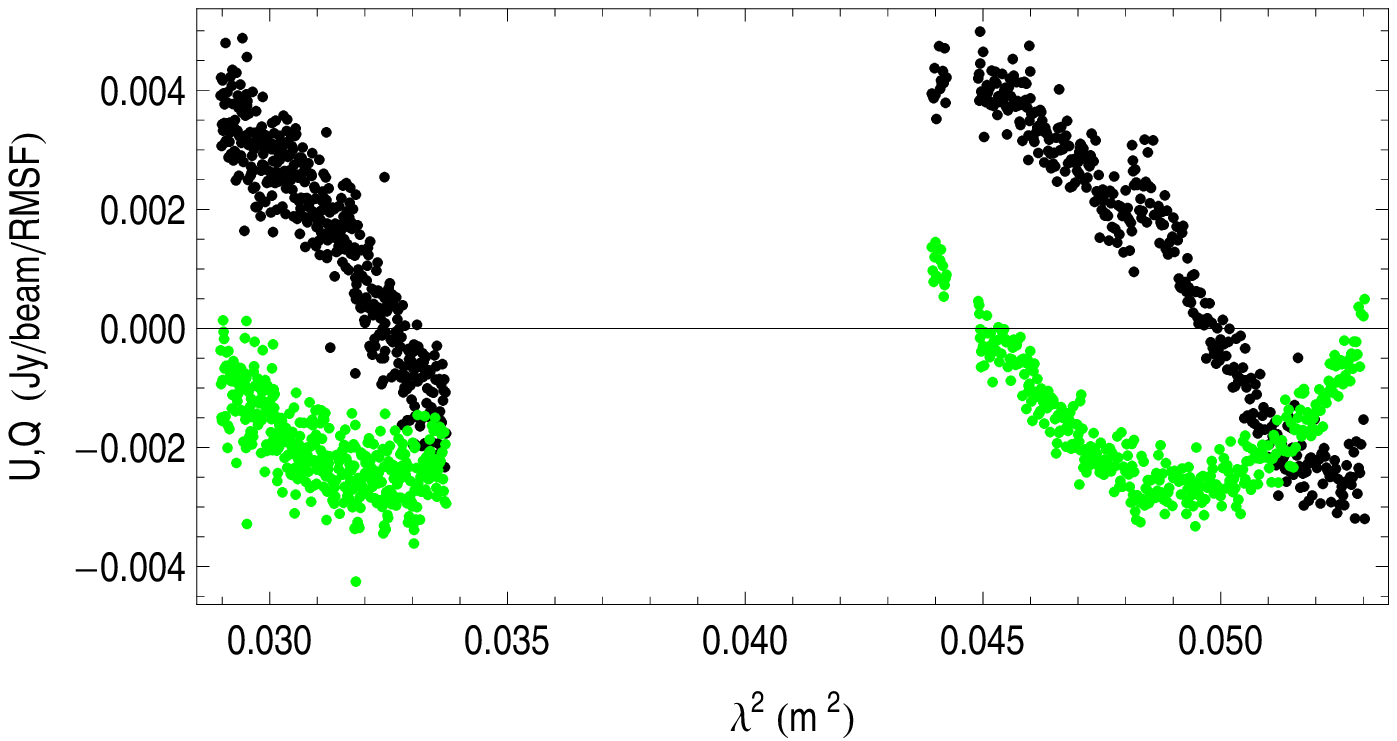}}
\put(0,110){\includegraphics[width=0.4\textwidth]{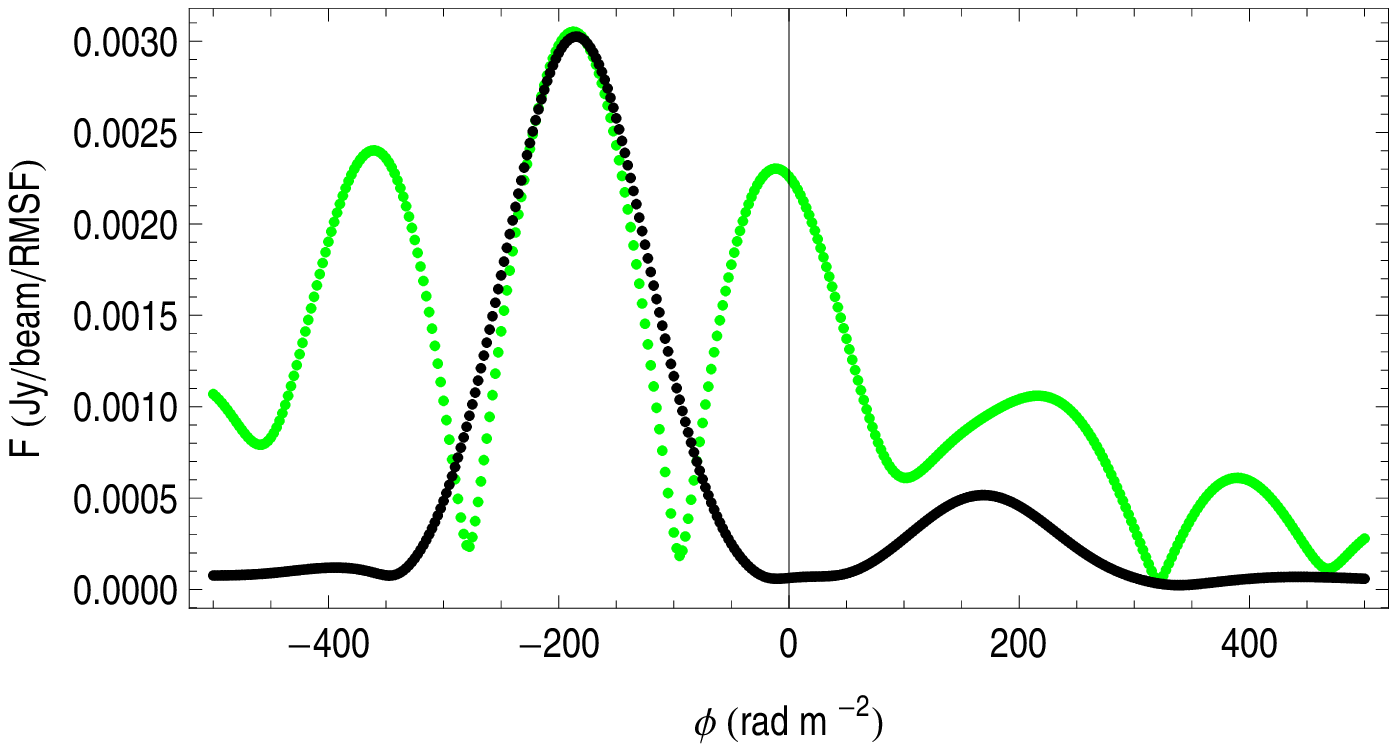}}
\put(5,0){\includegraphics[width=0.4\textwidth]{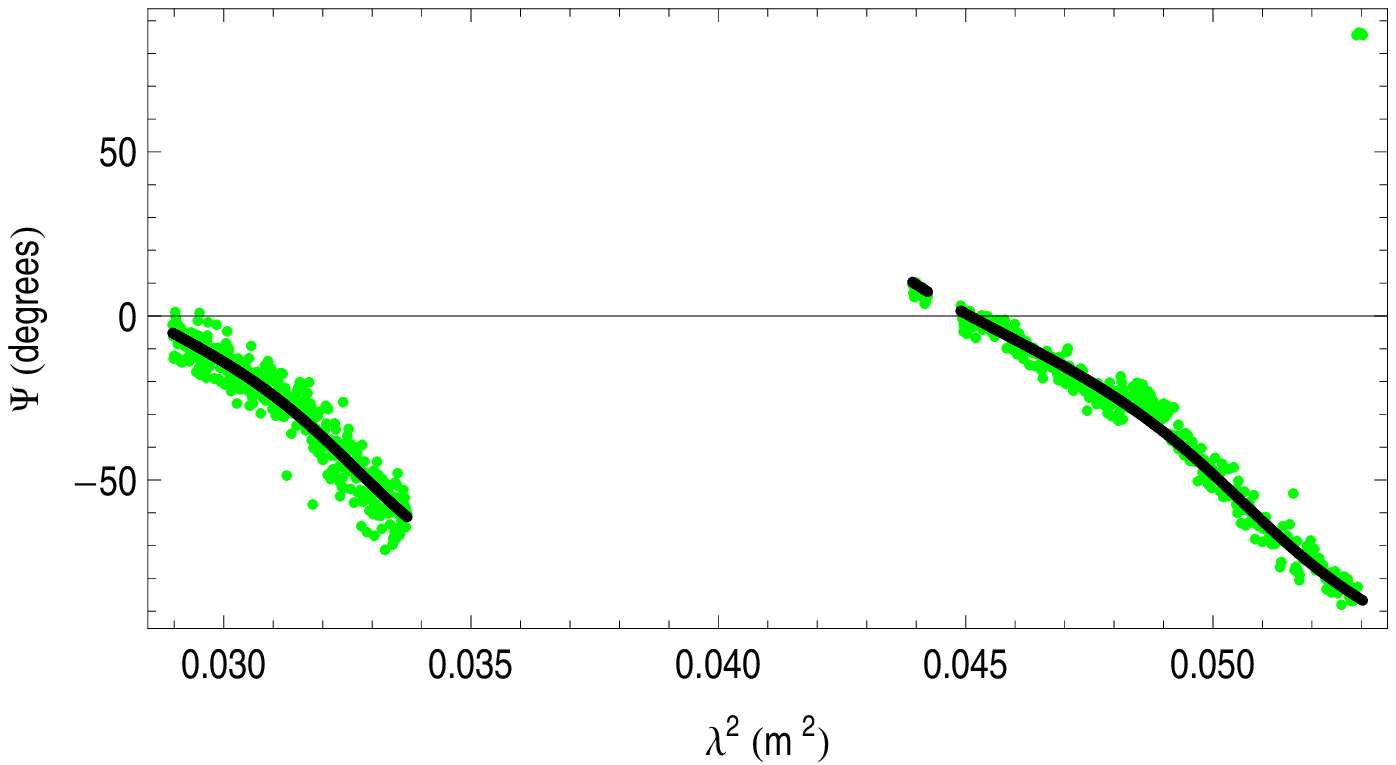}}
\put(210,320){(a)}
\put(210,210){(b)}
\put(210,100){(c)}
\end{picture}
\caption {Example of the RM-CLEAN process from
\protect\cite{2009IAUS..259..591H} for a bright point source in the field of
the galaxy NGC~7331. Panel (a): observed $Q$ (black) and $U$
(green); panel (b): the restored $F(\phi)$ before cleaning (green)
and after cleaning (black); panel (c): observed polarization angles
(green) and reconstructed after cleaning (black).} \label{fig_clean}
\end{figure}
\begin{figure}
\begin{picture}(240,355)(0,0)
\put(1,235){\includegraphics[width=0.4\textwidth]{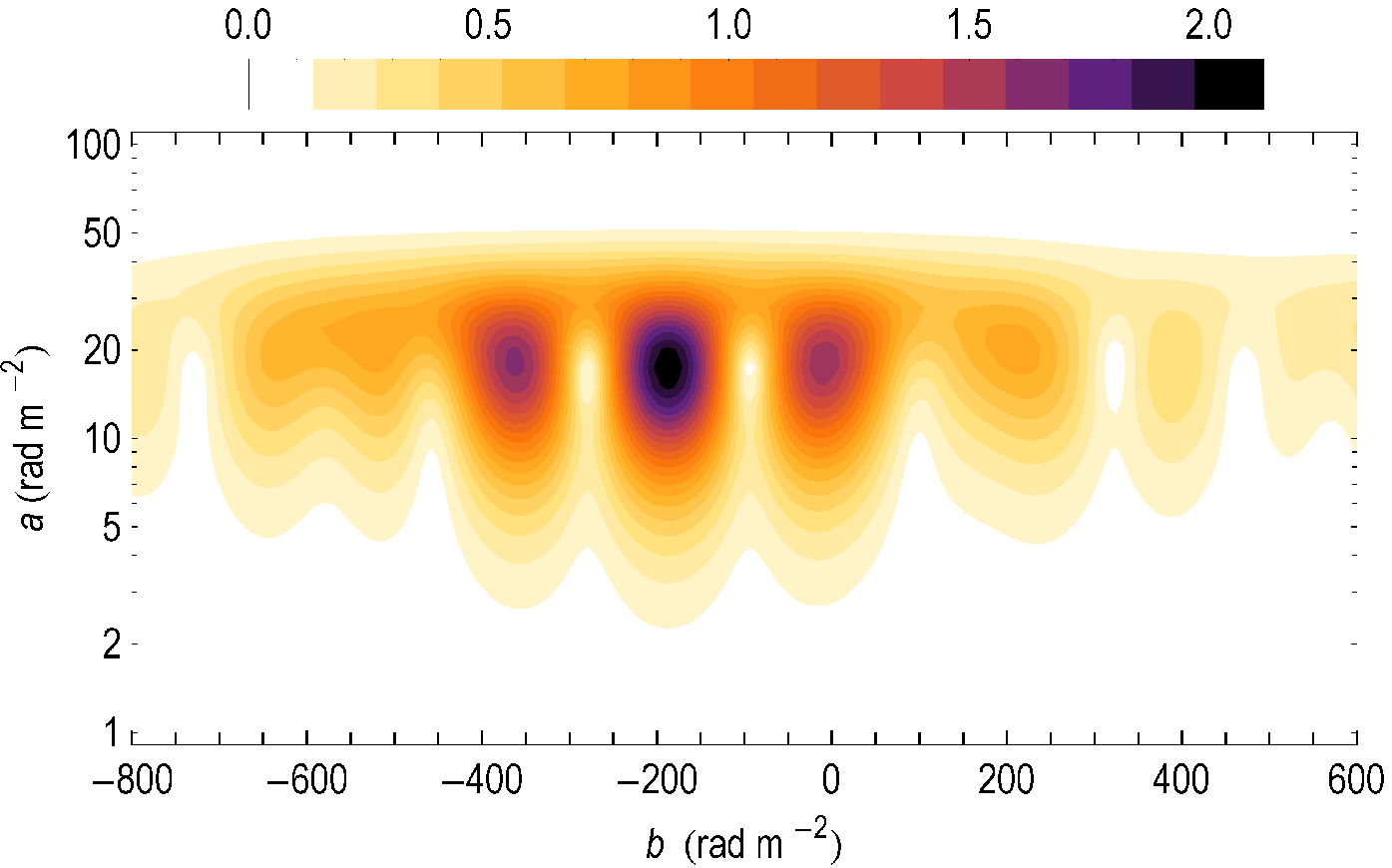}}
\put(1,110){\includegraphics[width=0.4\textwidth]{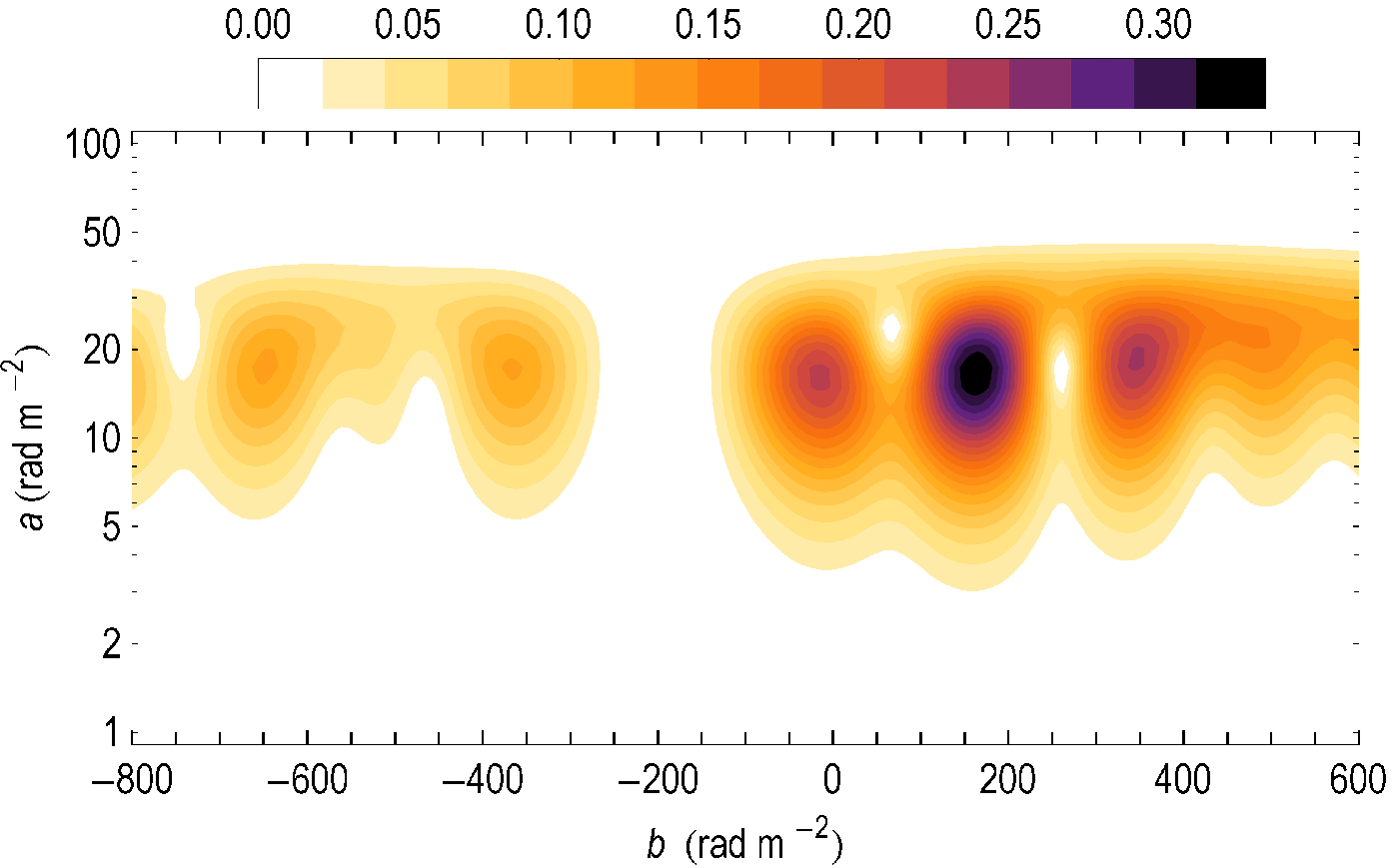}}
\put(0,00){\includegraphics[width=0.4\textwidth]{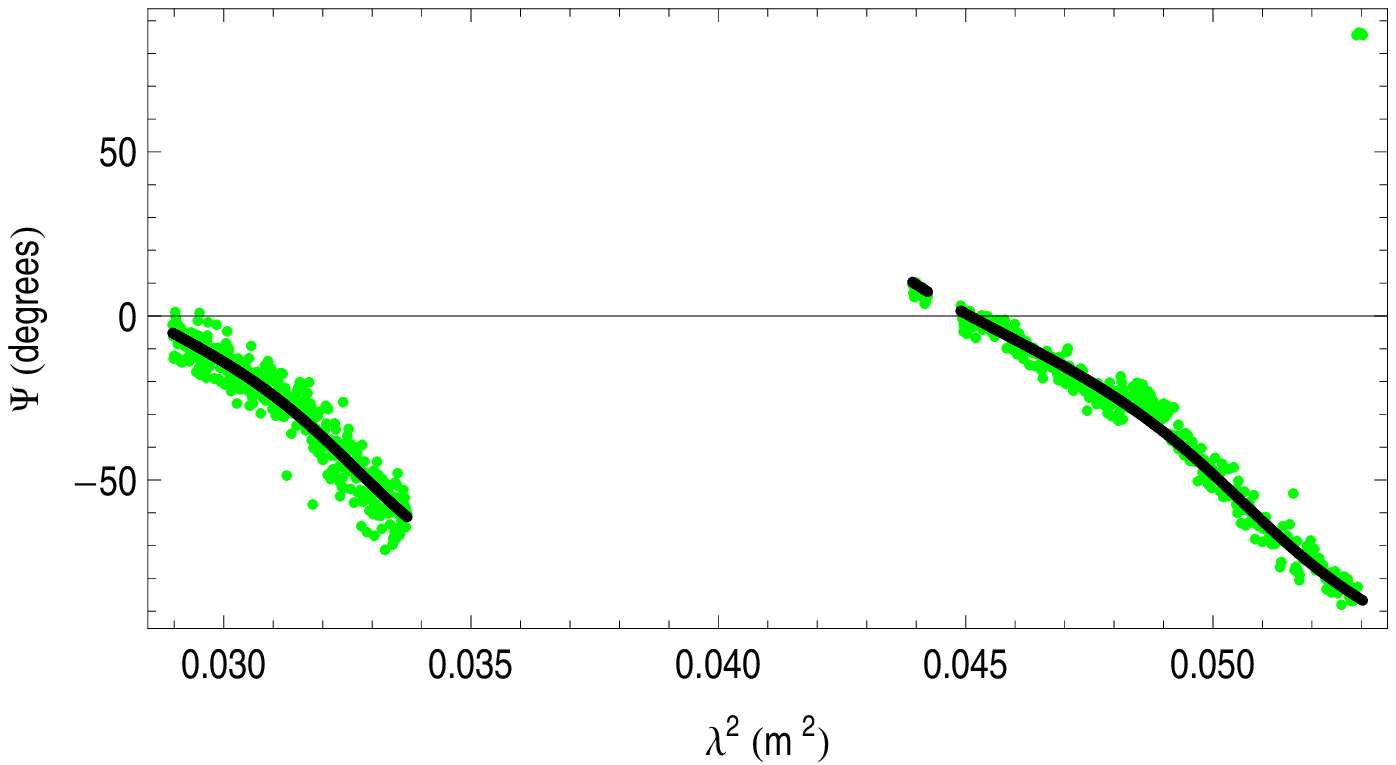}}
\put(210,340){(a)}
\put(210,220){(b)}
\put(210,100){(c)}
\end{picture}
\caption {Wavelet cleaning for the same example as in
Fig.~\ref{fig_clean}. Panel (a): wavelet plane after the first step
(the symmetry of the main maximum is used); panel (b): wavelet plane
after cleaning of the first point source and using symmetry argument
for the rest (note that the amplitudes in the two figures are
different); panel (c): observed polarization angles (green) and
reconstructed after two-step wavelet cleaning (black). Here and
below colors are given in a relative scale specific for each panel
and black color corresponds to maximum.} \label{fig_clean2}
\end{figure}

Let us show how this cleaning of objects, compact in Faraday depth space, can be realized in terms of
wavelet-based RM Synthesis. We use the example, presented in
Fig.~3 from \cite{2009IAUS..259..591H}. This figure is partially reproduced in
our Fig.~\ref{fig_clean}, where the observed $U(\lambda^2)$ and
$Q(\lambda^2)$, the restored $F(\phi)$ before and after cleaning and
the polarization angles observed and reconstructed after cleaning
are shown. The observations are done in two windows. In
Fig.~\ref{fig_clean2}a we show the modulus of the wavelet
coefficients $|w_{+}|$, calculated from same observed $U(\lambda^2)$
and $Q(\lambda^2)$. Then using the symmetry argument for the main
maximum we find $|w_{-}|$ and reconstruct $F(\phi)$. Taking the
dominating maximum in $F(\phi)$ at $\phi=\phi_0=-185$ rad m$^{-2}$ we
replace it by an equivalent point source and calculate for this
point source the wavelet transform $w_1$, based on the same sample
of observational channels as in original data. Then we subtract
$w_1$ from the wavelet transform $w$ of original data and we obtain
the wavelet image, shown in the second panel of
Fig.~\ref{fig_clean2}b, which demonstrates a dominating maximum at
$\phi=165$ rad m$^{-2}$. Using again the symmetry argument we get
$w_{-}$ and restore the second point-like source in $F(\phi)$. The
corresponding polarization angles, calculated for the given
observational windows are shown in the panel (c), which is
practically equivalent to the result of RM-CLEAN
(Fig.~\ref{fig_clean}). Summarizing this section, we conclude that,
in contrast to the extended sources, for a source that is point-like
in Faraday space, wavelet-based RM Synthesis gives just the same
result as conventional RM Synthesis.

\section{Galactic magnetic field in Faraday depth space}
\label{galactic}

\begin{figure}
\begin{picture}(240,220)(0,0)
\put(0,110){\includegraphics[width=0.4\textwidth]{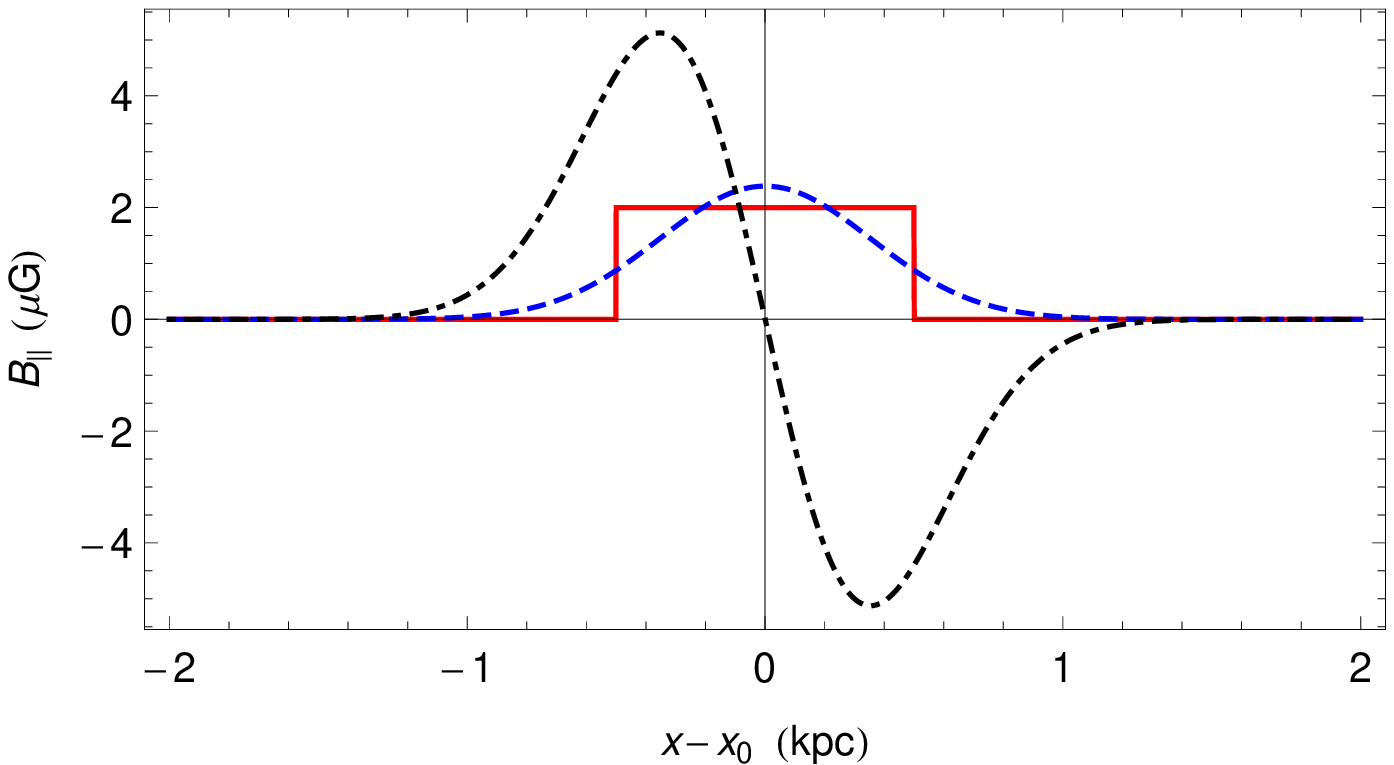}}
\put(0,00){\includegraphics[width=0.4\textwidth]{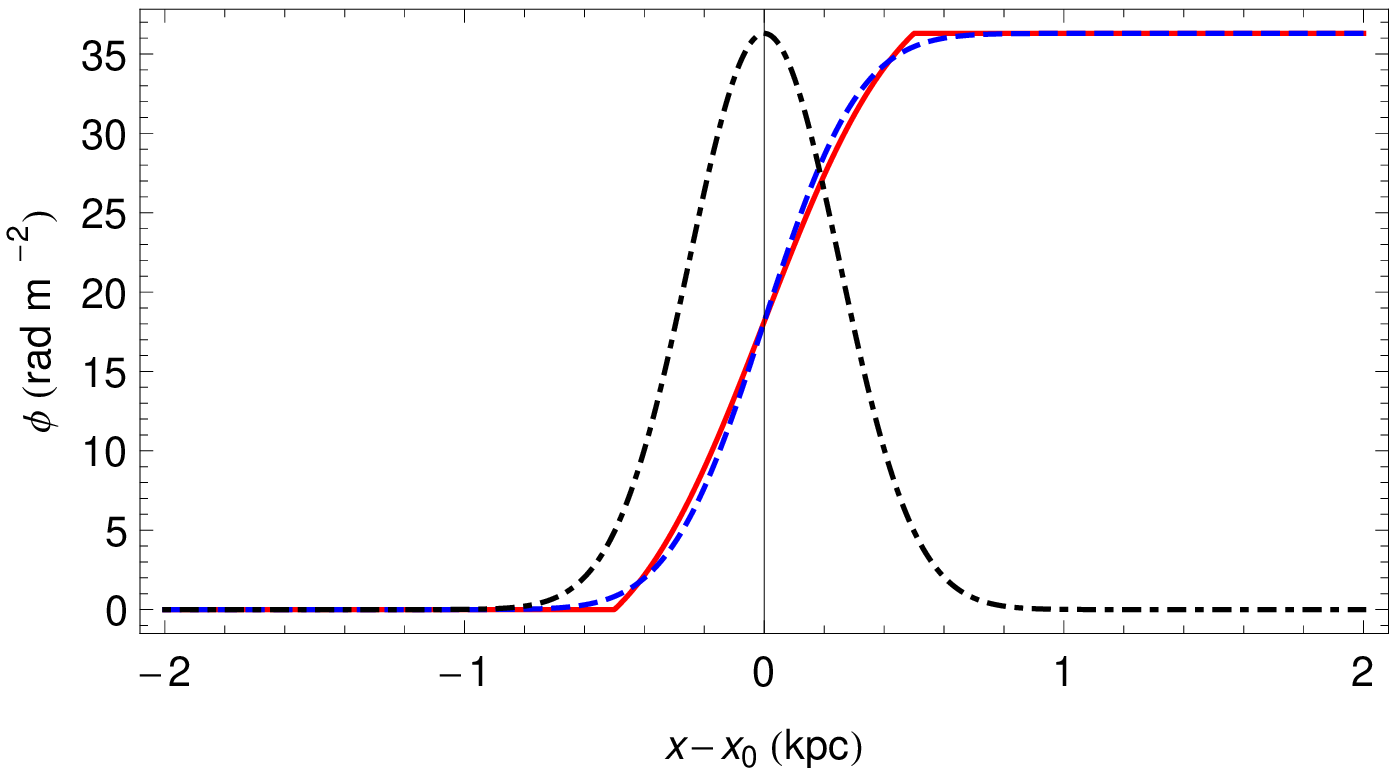}}
\put(210,210){(a)}
\put(210,100){(b)}
\end{picture}
\caption {Typical shapes of magnetic field distribution. Panel (a):
$B_1$ (solid line, red), $B_2$ (dashed, blue), $B_3$ (dot-dashed,
black); panel (b): corresponding Faraday depths $\phi$.  }
\label{fig1}
\end{figure}

The complex-valued intensity of polarized radio emission for a given
wavelength is
\begin{equation}
\label{p-def} P(\lambda^2) =  \int_{0}^{\infty}
\varepsilon(x){\rm e}^{2{\rm i}\chi(x)}{\rm e}^{2{\rm i}\phi(x) \lambda^2}  \mathrm{d} x,
\end{equation}
defined by the emissivity $\varepsilon$ and the intrinsic
polarization angle $\chi$ along the line of sight. The integral is
taken over the whole emitting region. The emissivity depends on the
relativistic (cosmic-ray) electron density $n_c$, the magnetic field
component perpendicular to the line of sight and the synchrotron
spectral index $\alpha$ (assuming $\alpha = 0.9$ below) as
(e.g. \cite{1965ARA&A...3..297G})
\begin{equation}
\label{vsreps} \varepsilon(x) = n_{\rm c}(x) |B_\perp(x) |^{1+\alpha} \, .
\end{equation}

\begin{figure}
\begin{picture}(240,220)(0,0)
\put(1,110){\includegraphics[width=0.4\textwidth]{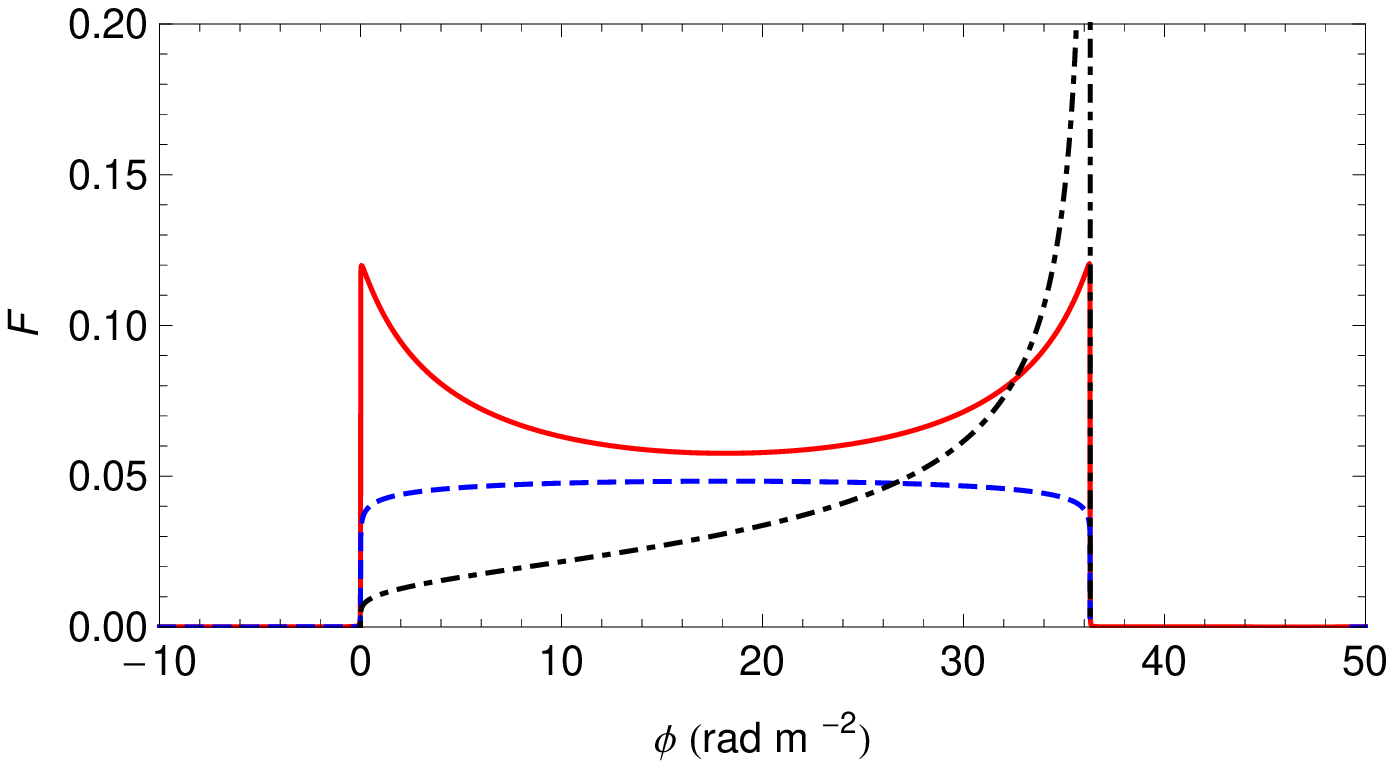}}
\put(0,00){\includegraphics[width=0.4\textwidth]{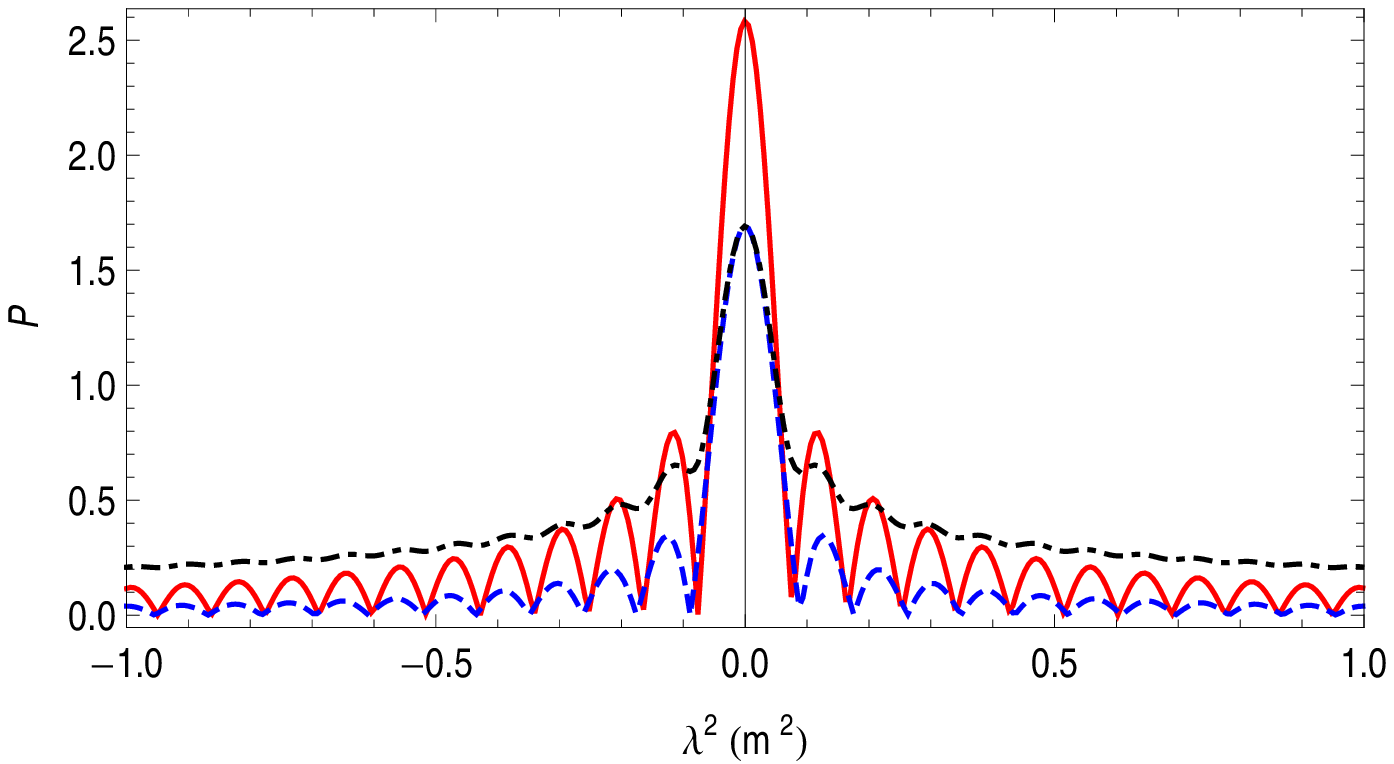}}
\put(210,210){(a)}
\put(210,100){(b)}
\end{picture}
\caption{Real part of the Faraday dispersion function $F(\phi)$
(panel a), ${\rm Im} \, F=0$, and the modulus of corresponding
$P(\lambda^2)$ (panel b) for different magnetic field profiles:
$B_1$ (solid line, red), $B_2$ (dashed, blue), $B_3$ (dot-dashed,
black). }
 \label{fig2}
\end{figure}

Our analysis starts from an investigation of the typical shapes of
the Faraday dispersion functions expected for lines of sight
crossing galactic discs with several simple distributions of the
parallel magnetic field, which provides the Faraday rotation. These
distributions are shown in Fig.~\ref{fig1}a. We consider two
symmetric distributions with respect to center at $x=x_0$. The first
one has a sharp boundary at $x=x_0\pm h$,
\begin{eqnarray}
\label{dis1}
 B_{1\parallel}(x) =
\left\{
  \begin{array}{ll}
   a_1, & |x-x_0|\leq h \\
   0, & |x-x_0|>h
  \end{array}.
\right.
\end{eqnarray}
Here $x_0$ is position of galactic equatorial plane and $h$ is the
half thickness of the galactic disc. The second distribution has a
smooth shape approximated by a Gaussian function with
 the scale height $h$,
\begin{equation}\label{dis2}
    B_{2\parallel}(x) = a_2 {\rm e}^{ -(x-x_0)^2/ h^2}.
\end{equation}

\begin{figure}\centering{
\includegraphics[width=0.4\textwidth]{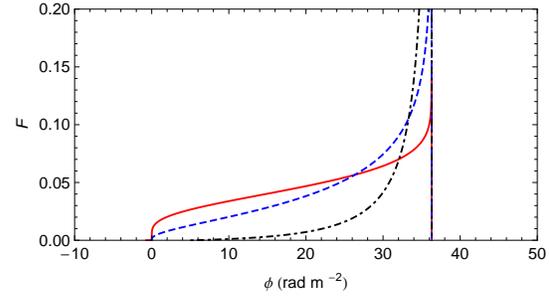}}
\caption {Faraday dispersion function $F(\phi)$ for different
separations $\Delta x$ between the regions of synchrotron emission
and Faraday rotation with Gaussian profiles: $\Delta x =0.1$ kpc
(solid line, red), $\Delta x=0.2$ kpc (dashed, blue), $\Delta x=0.5$ kpc
(dot-dashed, black).}
 \label{fig2a}
\end{figure}

For comparison we take into consideration a distribution with a
field reversal along the line of sight
\begin{equation}\label{dis3}
    B_{3\parallel}(x)=-a_3{{(x-x_0)} \over {h}}{\rm e}^{ -(x-x_0)^2/ h^2}.
\end{equation}
The amplitudes $a_2$ and $a_3$ are adjusted to obtain the same
maximal Faraday depth for all three distributions.
We choose an amplitude of the magnetic field of $a_1=2\,\mu{\rm G}$
and $h=0.5\,{\rm kpc}$. The densities of relativistic electrons
$n_{\rm c}$ (responsible for synchrotron emission) and thermal electrons
$n_{\rm e}$ (responsible for Faraday rotation) measured in cm$^{-3}$ are
assumed to have Gaussian profiles
\begin{equation}\label{dise}
    n_{\rm c}(x)= C {\rm e}^{ -(x-x_0)^2/ h_{\rm c}^2}, \hspace{1cm} n_{\rm e}(x)= 0.03\, {\rm e}^{ -(x-x_0)^2/ h_{\rm e}^2}.
\end{equation}
Here $C$ (measured in cm$^{-3}$) is the relativistic electron
density at the galactic midplane. We assume that the scale height
of the relativistic electrons $h_{\rm c}$ is twice that of the thermal
electrons $h_{\rm e}$ and that the scale height of the magnetic field $h$
is the same as that of the thermal electrons $h_{\rm e}$. We use $C$ as
a normalization constant for $F$ and $P$, so that they are
numerically evaluated in arbitrary but mutually consistent units.
 Fig.~\ref{fig1}b
shows that the symmetric profiles of the parallel magnetic field
(first two cases) lead to a monotonic increase (decrease) of Faraday
depth along the coordinate $x$ inside the galaxy, while the
antisymmetric profile (last case) leads to  an  inversion - the
central galactic plane becomes the most distant (or most nearby)
point in Faraday depth space.

\begin{figure}
\begin{picture}(240,500)(0,0)
\put(0,375){\includegraphics[width=0.40\textwidth]{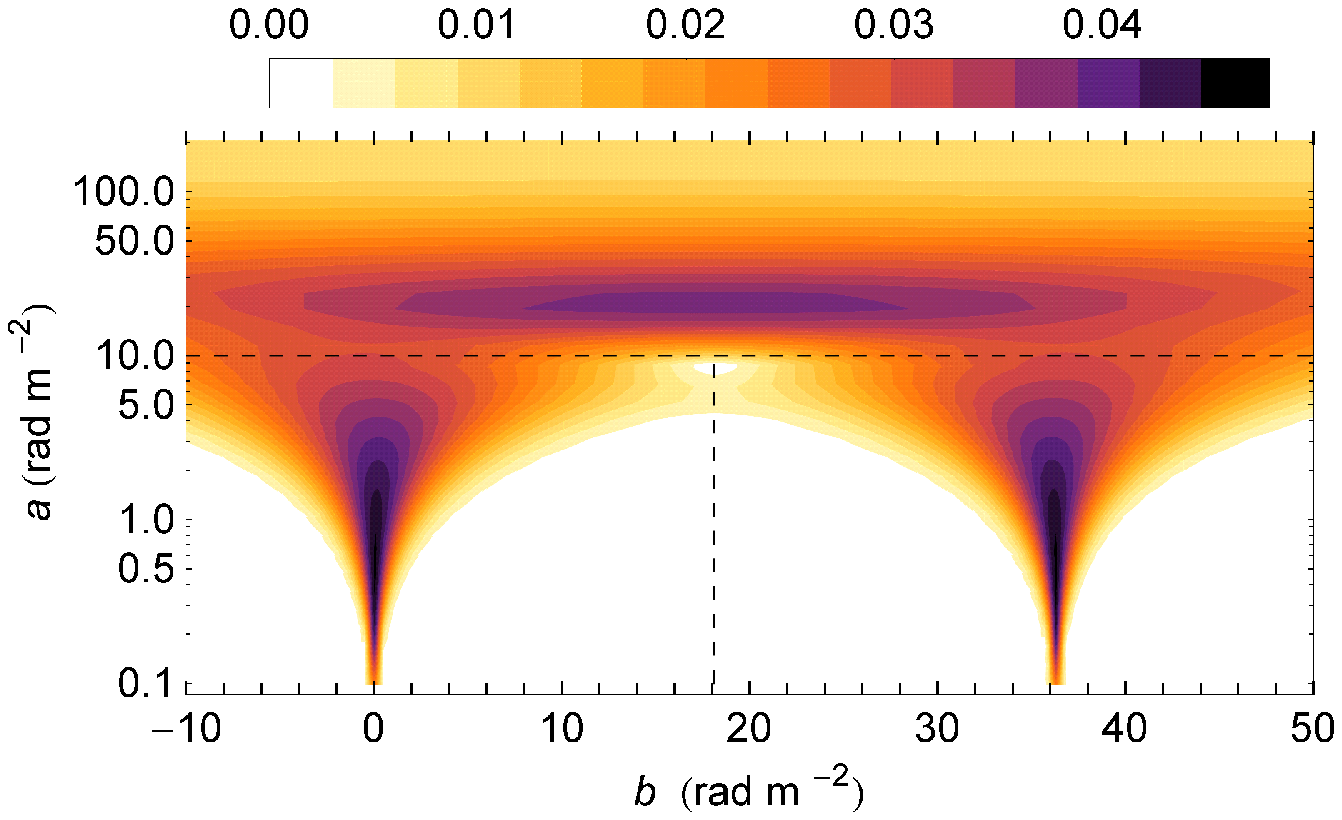}}
\put(0,250){\includegraphics[width=0.40\textwidth]{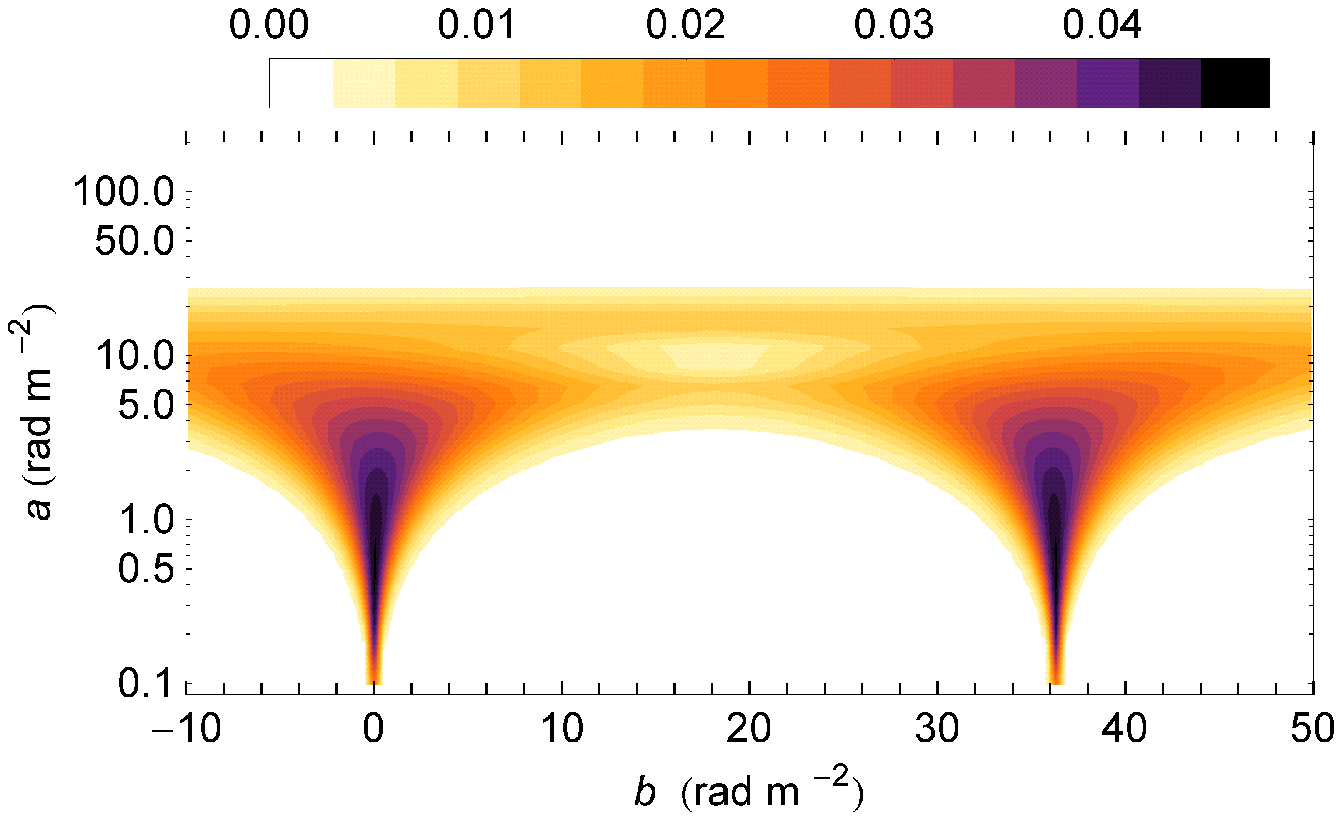}}
\put(0,125){\includegraphics[width=0.40\textwidth]{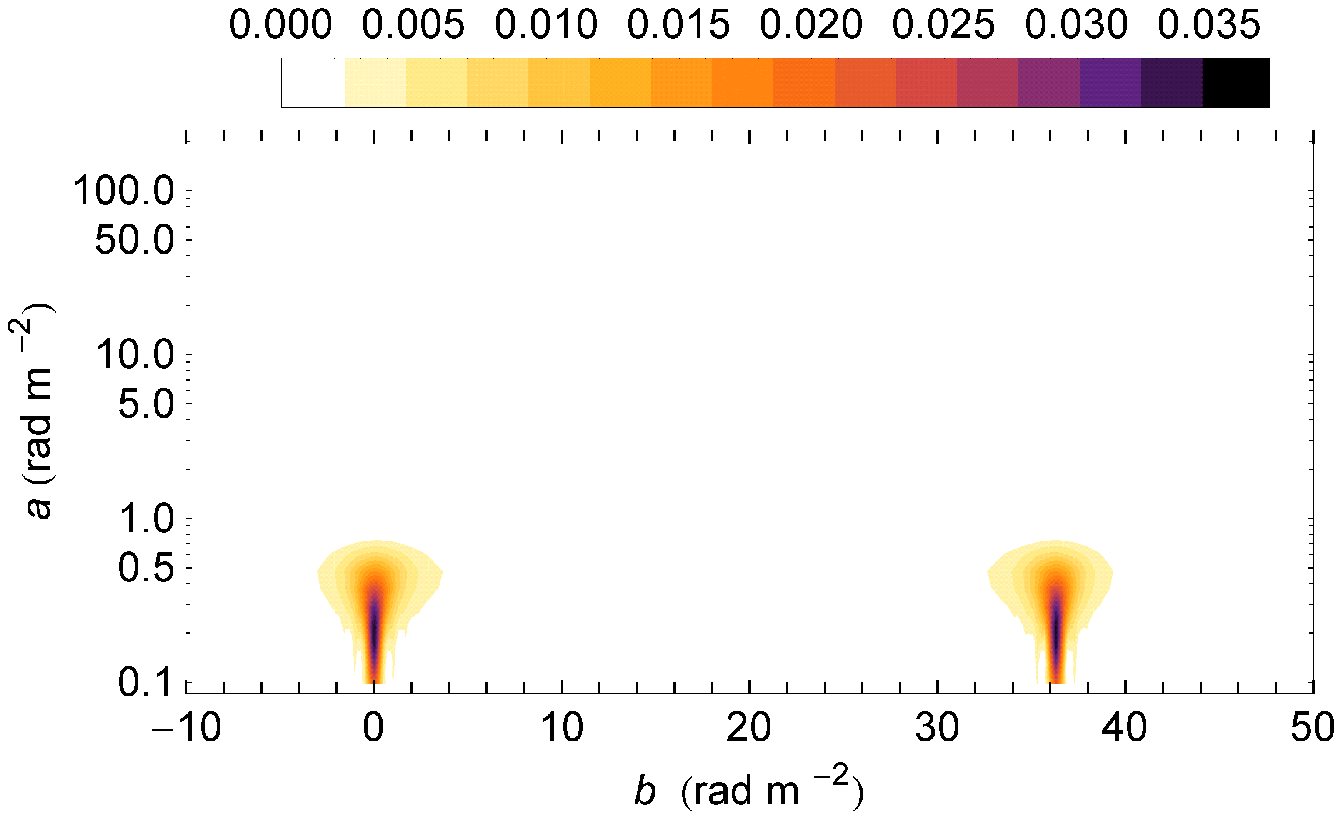}}
\put(0,000){\includegraphics[width=0.40\textwidth]{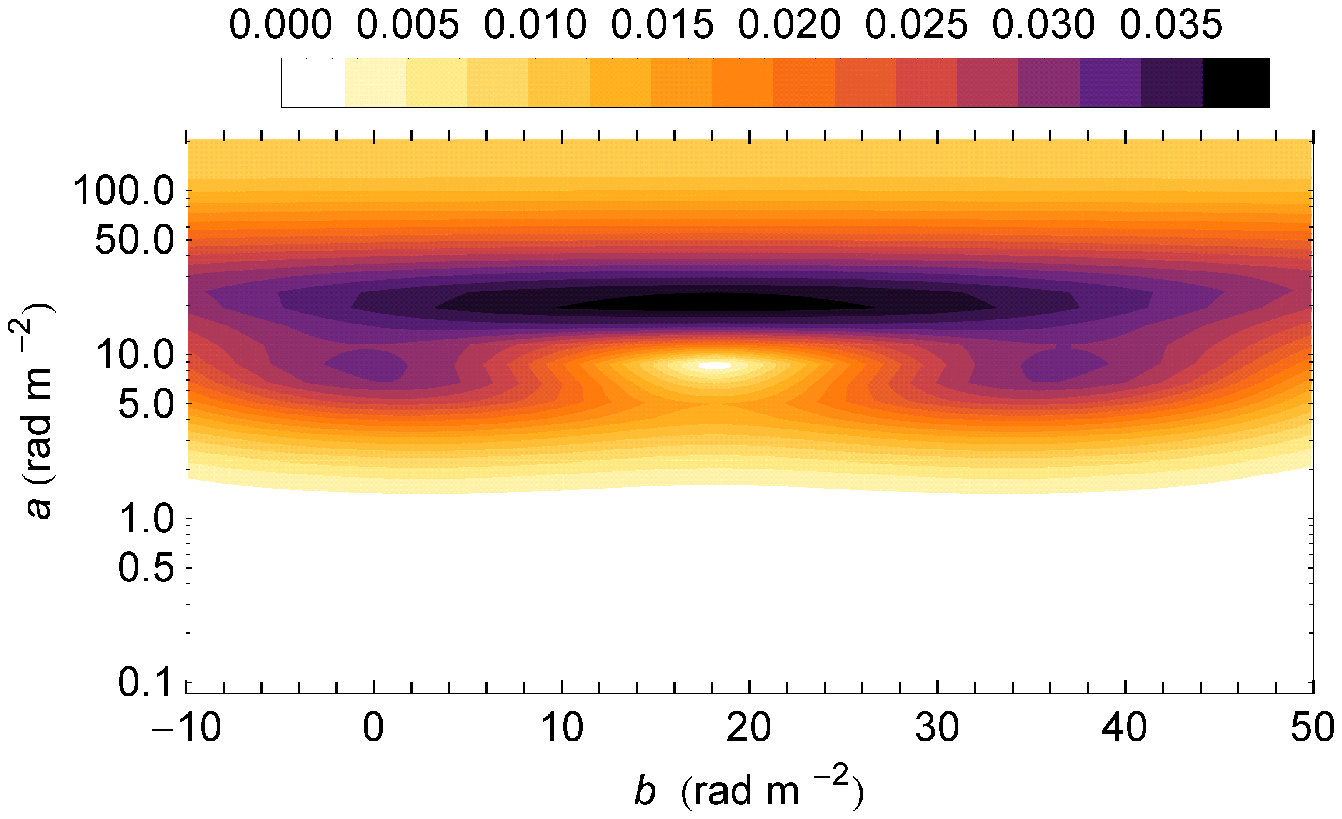}}
\put(210,485){(a)}
\put(210,360){(b)}
\put(210,235){(c)}
\put(210,110){(d)}
\end{picture}
\caption {Wavelet plane $|w_F(a,b)|$ of the example of
Figs.~\ref{fig1} and \ref{fig2} (solid lines)  for various spectral
windows. Panel (a): $0.06<\lambda<2.5$\, m; panel (b):
$0.21<\lambda<2.5$\, m; panel (c): $1.25<\lambda<2.5$\, m; panel
(d): $0.06<\lambda<0.21$\,m.  } \label{fig6}
\end{figure}

For the sake of normalization we choose the direction of the
perpendicular field to obtain a purely real $P$ (i.e. zero
polarization angle).
The perpendicular component is described by the same distributions
as for the parallel component in Eqs.~(\ref{dis1})-(\ref{dis3}).
Namely, we take $B_{1\perp}=B_{1\parallel}$,
$B_{2\perp}=B_{2\parallel}$ and $B_{3\perp}=B_{2\parallel}$.

Fig.~\ref{fig2}a shows that both even profiles ($B_1$ and $B_2$)
produce in Faraday space basically box-like distributions with sharp
boundaries. In contrast, Faraday depth in the case of the magnetic
field with a reversal varies slowly near the point where Faraday
depth becomes minimal and has a sharp point-like structure in
$F(\phi)$ near the point where $\phi$ is maximal while $F$ is still
nonvanishing,  in the given examples $F$ is even maximal at
this point.

\begin{figure}
\begin{picture}(240,220)(0,0)
\put(0,110){\includegraphics[width=0.4\textwidth]{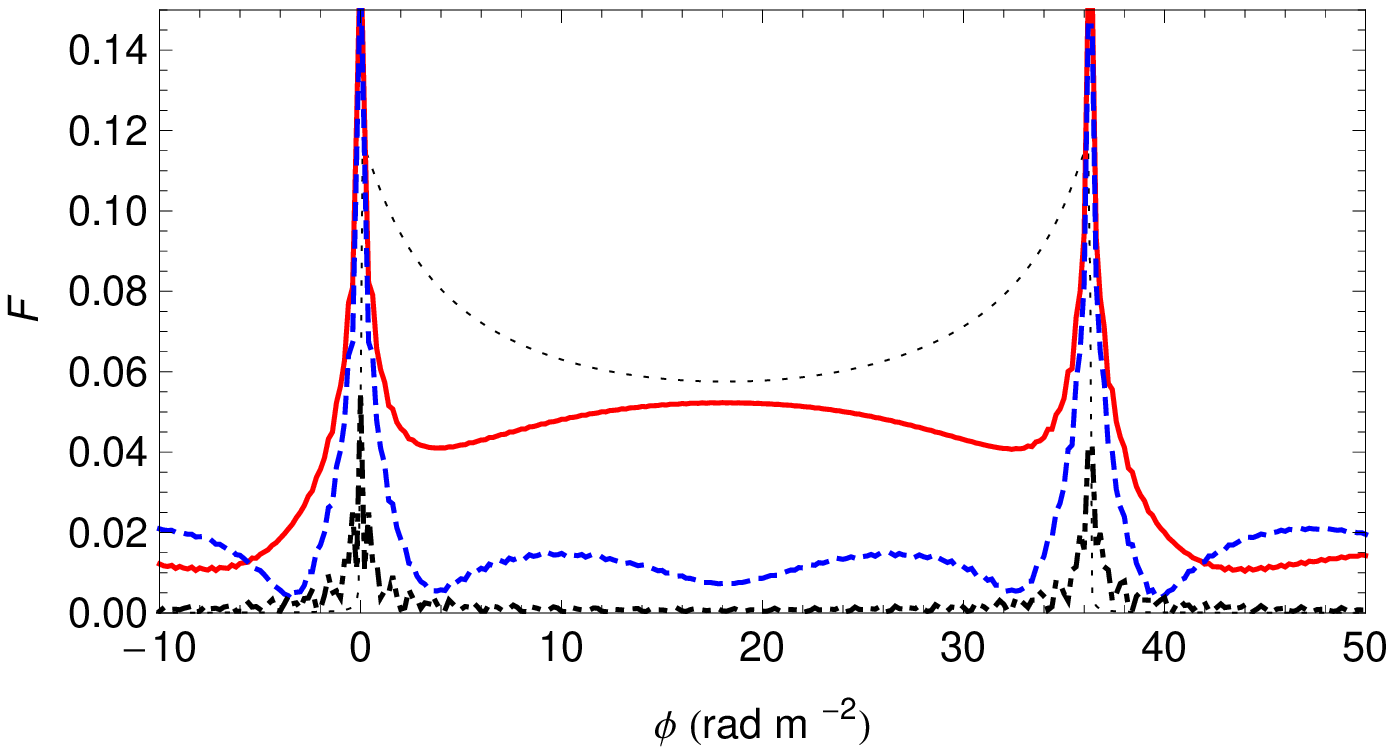}}
\put(0,00){\includegraphics[width=0.4\textwidth]{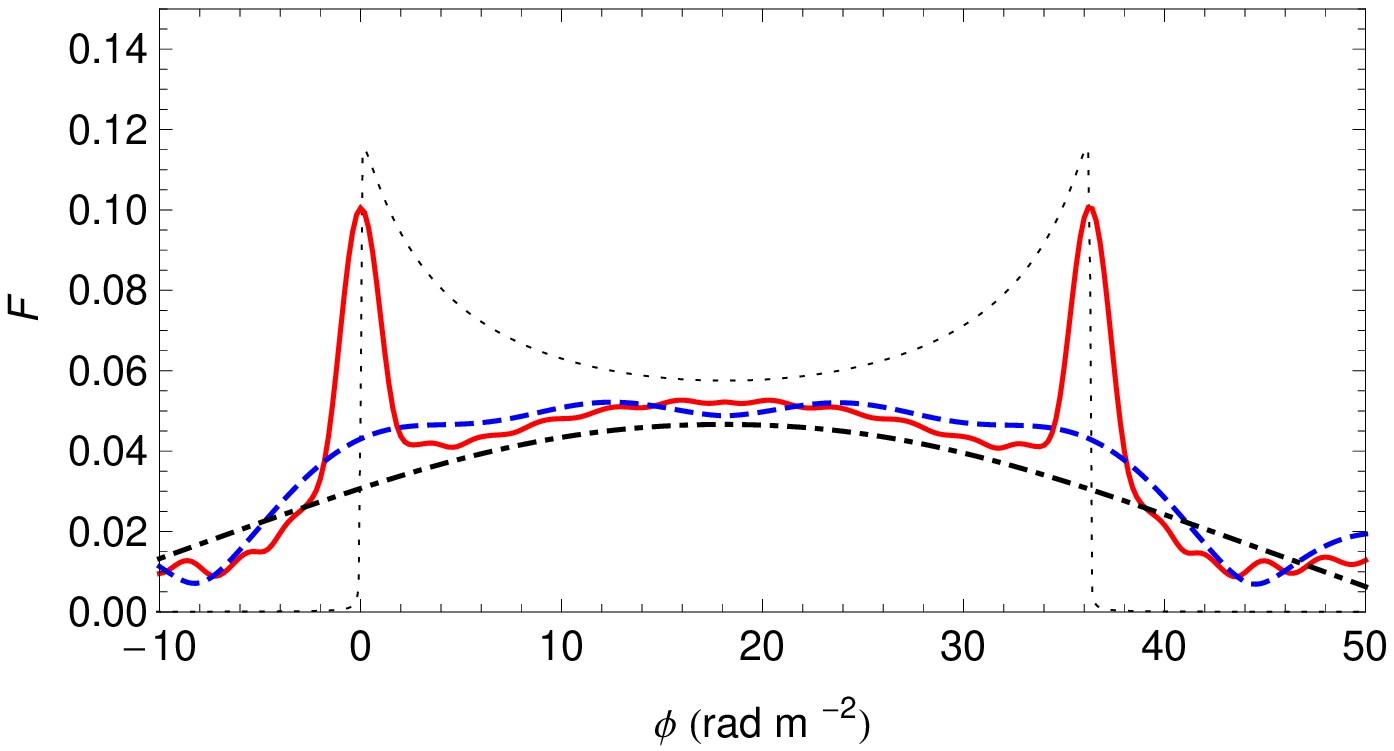}}
\put(210,210){(a)}
\put(210,100){(b)}
\end{picture}
\caption {Test Faraday dispersion function (black dots in both
panels), the same as shown in Fig.~\ref{fig2}a (solid red line), and
results of RM Synthesis performed from the wavelet planes shown in
Fig.~\ref{fig6} for various spectral windows. Panel (a):
$0.06<\lambda<2.5$\, m (solid, red line), $0.21<\lambda<2.5$\, m
(dashed, blue), $1.25<\lambda<2.5$\, m (dash-dot, black); panel (b):
$0.06<\lambda<1.0$\, m (solid, red line), $0.06<\lambda<0.42$\, m
(dashed, blue), $0.06<\lambda<0.21$\, m (dash-dot, black).}
\label{fig66}
\end{figure}

We conclude from Fig.~\ref{fig2}a that box-like and point-like
shapes of $F(\phi)$  are common in  simple models of
disc components of spiral galaxies. Of course, one can obtain a
smoother distribution of $F(\phi)$ by taking a wider distribution
for thermal electrons than for relativistic ones. According to
Fig.~\ref{fig2}b the level of polarized intensity at large
$\lambda^2$ is higher for pointed $F$ profiles (hence easier to
recognize by low-frequency observations) than for box-like ones.

In general, the synchrotron emission is not necessarily located in
the same region as the Faraday rotation. This leads to a so-called
``Faraday screen'' which can be due to a specific magnetic field
configuration (strong $B_{\parallel}$, weak  $B_{\perp}$) or to a
difference between distributions of $n_{\rm c}(x)$ and $n_{\rm e}(x)$. For
simplicity, we shift $\varepsilon(x)$ by some distance $\Delta x$.
The resulting Faraday dispersion functions for Gaussian shape (the case $B_2$ in Fig.~\ref{fig1})
and different Faraday
screen separations are shown in Fig.~\ref{fig2a}. For larger values
of $\Delta x$ the Faraday dispersion function $F$ becomes more
pointed and asymmetric.

\section{The role of the spectral range}
\label{range}

The efficiency of RM Synthesis crucially depends on the
observational range $\lambda_{\rm min} < \lambda < \lambda_{\rm
max}$ \citep{2005A&A...441.1217B}.
Wavelets give a helpful illustration for the role of these parameters.

Fig.~\ref{fig6} shows the wavelet planes for the first artificial
example (solid lines in Fig.~\ref{fig2}), calculated for different
observational windows. Each panel shows the absolute value of the
wavelet coefficient $W(a,b)$ in the $(a,b)$ plane. The horizontal
axis ($b$) presents the position in Faraday depth space, the
vertical axis gives the scale $a$ in logarithmic presentation. The
upper bound is first fixed at $\lambda_{\rm max}=2.5$\,m. The panel
(a) shows the wavelet plane simulated for the lower bound
$\lambda_{\rm min}=6$\,cm (thus $0.0036<\lambda^2<6.25$\, m$^2$).
The large horizontal dark feature at $a \approx 40$\, rad m$^{-2}$
corresponds to the box-like structure, and the two tapering
structures are generated by the sharp borders of this box. The next
panel shows what remains in the wavelet plane if the lower bound is
moved to $\lambda_{\rm min}=21$\,cm (the window
$0.044<\lambda^2<6.25$\, m$^2$). The central spot (responsible for
the box) is almost completely lost even at these relatively short
wavelengths. Applying the actual LOFAR highband window
($1.56<\lambda^2<6.25$\, m$^2$), only weak traces of side horns
remain in the wavelet plane (panel c). The lower panel (d) in
Fig.~\ref{fig6} shows the wavelet plane for a relatively
high-frequency window ($0.06<\lambda<0.21$\, m), which perfectly
keeps all information concerning the large scales, but completely
loses any information concerning the small scales, responsible in
this example for the abrupt boundaries.


 For applying the symmetry argument expressed by Eq.~(\ref{w-w+}) the position of each local maximum $\phi_0$ and
the domain of its influence should be defined. In Fig.~\ref{fig6}a we recognize three maxima
and we divide the wavelet plane in three parts: one domain corresponds $a>10$ rad m$^{-2}$  and two domains for $a\leq10$ rad m$^{-2}$  are separated at $b\approx18$ rad m$^{-2}$. Separatrices are shown in
Fig.~\ref{fig6}a by dashed lines. In the case shown in Fig.~\ref{fig6}c only two lower domains can be recognized. Note, that two horns-like structures are associated with the sharp borders, which are not symmetric objects,  and the application of the symmetry assumption to them does not improve the reconstruction.


Fig.~\ref{fig66} shows how the shrinking of the wavelength range of
observations reduces the quality of the Faraday function
reconstructed from the corresponding sets of $w_F(a,b)$. We see that
increasing of $\lambda_{\rm min}$ leads to a
gradual decay of the reconstructed $F$ in the main bulk of
the structure (Fig.~\ref{fig66}a). Starting
from $\lambda_{\rm min} = 21$ cm we reconstruct details associated
with the boundaries only.  Decreasing the upper bound $\lambda_{\rm
max}$ we lose the small-scale details, restoring a smooth structure
only (Fig.~\ref{fig66}b).

The possibility to reconstruct a structure of scale $\Delta \phi$ in
Faraday space is determined by the quantity $X= \Delta \phi \,
\lambda^2$ where $\lambda$ is the wavelength at which the
observations are performed. We illustrate the role of this quantity
for the large-scale structure from Fig.~\ref{fig66}a. The relevant
parameters here are $\Delta \phi = 38$ rad m$^{-2}$ and $\lambda =
\lambda_{\rm min}$. $X=1.68$ is for the dashed (blue) line and
$X=60$ for the dash-dot (black) lines. Because the latter curve
keeps the information of sharp ends of the distribution $F(\phi)$
and even the first curve looses a substantial part of the signal,
\begin{equation}
X = \Delta \phi \, \lambda_{\rm min}^2 \lesssim 1
 \label{crit}
\end{equation}
is the condition to observe a galaxy as an extended source in
Faraday space rather than as a couple of point-like sources
\citep{2005A&A...441.1217B}. For observations in the LOFAR highband ($\lambda_{\rm
min} = 1.25$\,m) this gives $\Delta \phi = 0.64$\,rad m$^{-2}$,
which is at least one order of magnitude lower than a typical value
for contemporary galaxies. If $\lambda_{\rm min} =0.167$\,m (by
including observations e.g. with the ASKAP telescope) this yields a
much more comfortable $\Delta \phi=35$\,rad m$^{-2}$.

For distant galaxies the wavelength in the framework of the source
is $\lambda/(1+z)$ ($z$ is the redshift) and Eq.~(\ref{crit}) reads
\begin{equation}
X = \Delta \phi \, \lambda_{\rm min}^2 /(1 +z)^2\lesssim 1.
 \label{zcrit}
\end{equation}
Taking $Z=3$ one obtains that $\Delta \phi \approx 10$\,rad m$^{-2}$
for LOFAR-type observations. LOFAR observations can also be used to
isolate a contribution of weak intergalactic magnetic fields, which
can give typical $\Delta \phi$ much lower than that for galactic
magnetic fields.

\begin{figure}
\begin{picture}(240,220)(0,0)
\put(0,110){\includegraphics[width=0.4\textwidth]{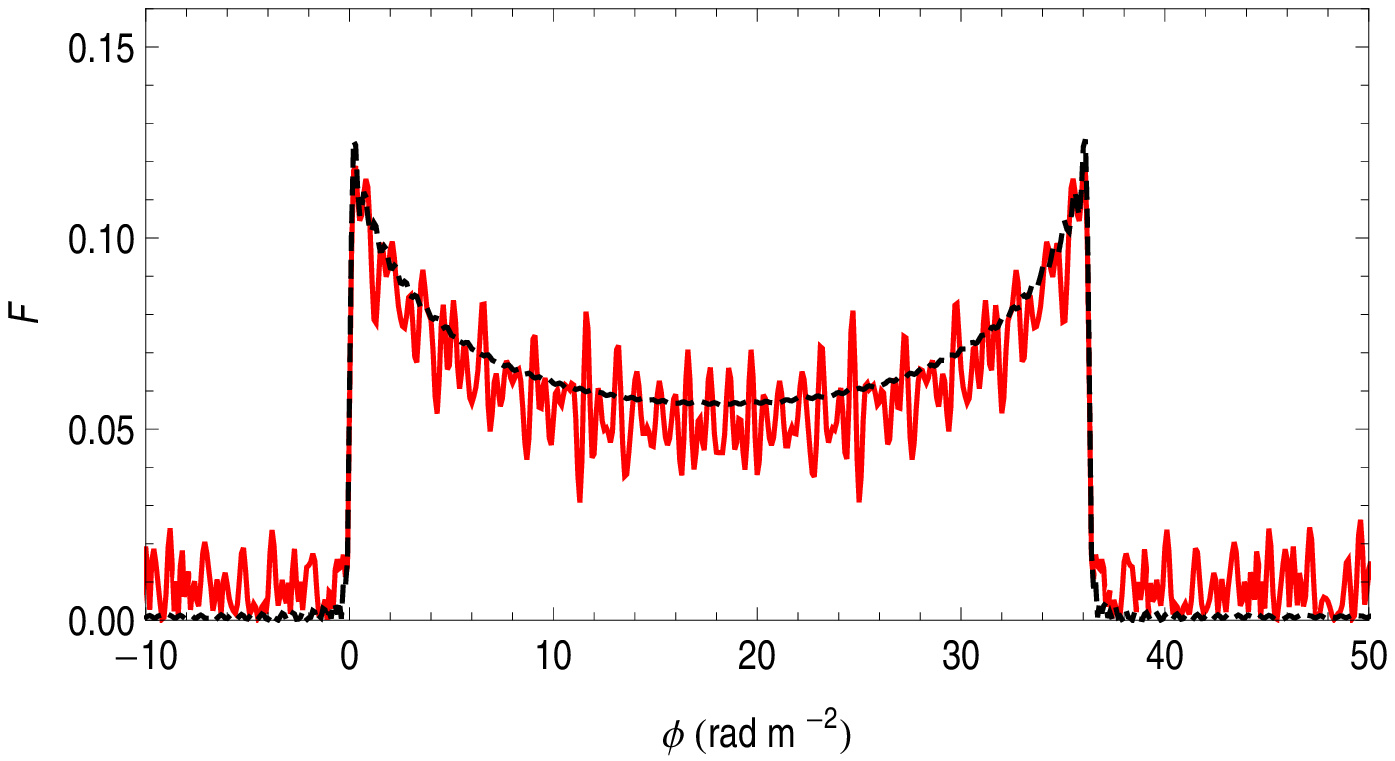}}
\put(0,00){\includegraphics[width=0.4\textwidth]{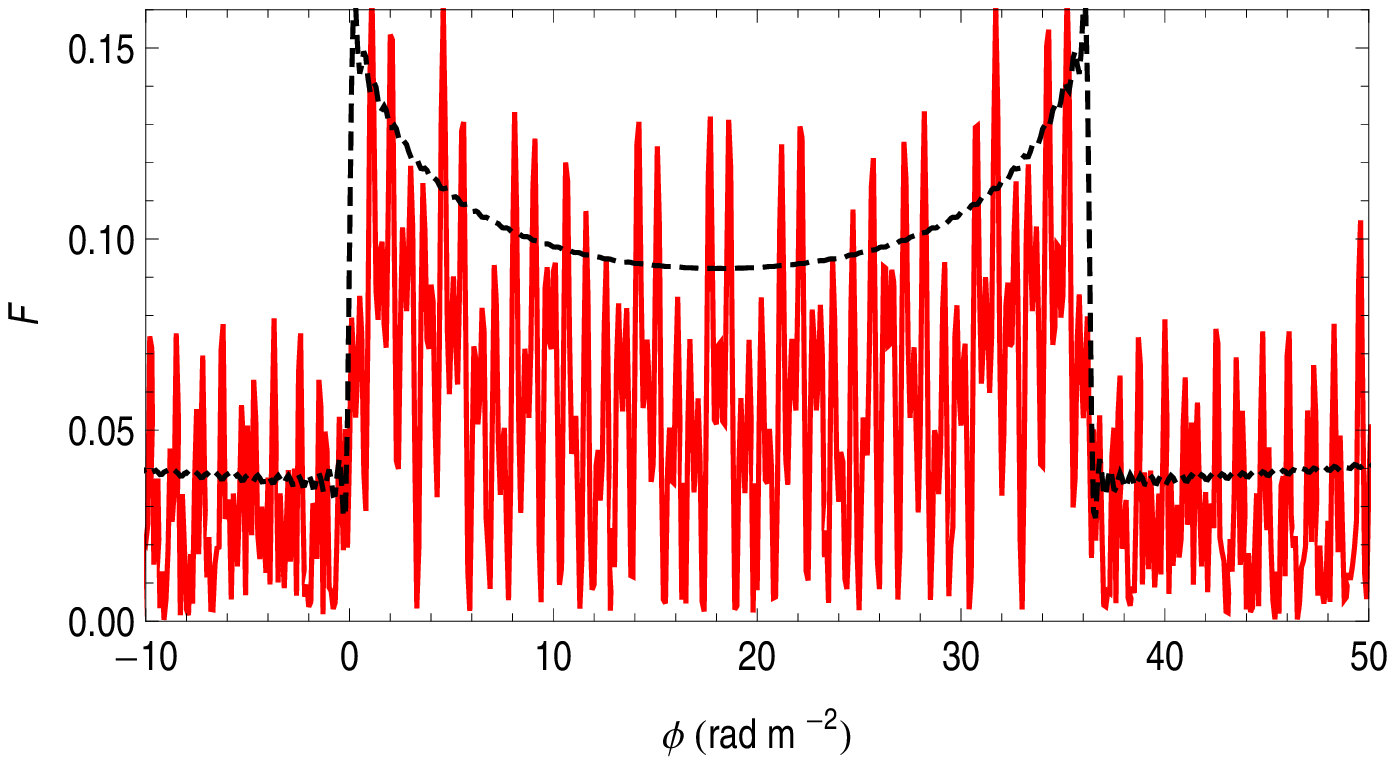}}
\put(210,210){(a)}
\put(210,100){(b)}
\end{picture}
\caption {Modulus of the Faraday dispersion function $F$ obtained
from wavelet-based RM Synthesis performed for the observational
window $0.06<\lambda<2.5$\, m  with 1024 (panel a) and 128 (panel b)
channels providing uniform sampling in $\lambda^2$ space (dashed
black line) or uniform sampling in frequency space (solid red line).
} \label{fig_sampling}
\end{figure}

As we conclude above, RM Synthesis of extended magnetic
configurations requires a proper wavelength range of observations.
$\lambda_{\rm min}$ has to be small enough to reproduce the grand
design of $F$ and a large $\lambda_{\rm max}$ is required to
reproduce sharp details of $F$. Obviously, RM Synthesis demands a
substantial number of channels in the observational range to perform
the Fourier transform included in the method. A dispersion
function $F$ which is extended over a large range of Faraday depths
requires a large number of channels. If the number of channels is
limited, the main question is how to distribute the channels
between the given values of $\lambda_{\rm min}$ and $\lambda_{\rm
max}$. We clarify this using the example  considered for
Fig.~\ref{fig6}. For the sake of the definiteness we compare two
samplings, i.e. a uniform spacing in frequency  space and that one
in the $\lambda^2$ space.

We presume that the spectral range of observations is sufficiently
large and use $\lambda_{\rm min} = 6$\,cm and $\lambda_{\rm max} =
2.5$\,m and apply the wavelet-based RM Synthesis.
Fig.~\ref{fig_sampling}a presents the results for a large (1024)
number of channels. We see that even here a uniform spacing in
$\lambda^2$ looks better than
one in frequency space, which
gives a more noisy result. The difference between the two kinds of
spacings becomes much more pronounced for a moderate (128) number of
channels. The results for the uniform spacing in $\lambda^2$ is more
or less similar to that one in the upper panel, while for a uniform
spacing in frequency the noise in $F$ becomes comparable to the
signal. It is clear that the effect is essential for large spectral
windows (namely, for a large ratio $\lambda_{\rm max}/\lambda_{\rm
min}$) because a uniform spacing in frequency leads to points
crowding at small $\lambda^2$. A poor sampling at large $\lambda^2$
(essential to detect small scales in Faraday depth space) results in
small-scale noise in the reconstructed signal. Note that the
advantages of uniform spacing in $\lambda^2$ was first discussed by
\cite{1979A&A....78....1R}.

\section{Small-scale magnetic fields}
\label{turb}

In contrast to the traditional methods, wavelet-based RM Synthesis
opens a new option to quantify the small-scale (turbulent)
component of magnetic field. Remember that it is widely
believed that one can isolate two contributions in total magnetic
field, i.e. the large-scale (regular) magnetic field and the
small-scale one. Correspondingly, the Faraday dispersion function
$F$ and the Faraday depth $\phi$ have two corresponding components,
coming from large-scale and turbulent fields. The traditional Burn
(1966) theory suggests that averaging $P$ over turbulent variations
yields an additional depolarization  term like internal Faraday
dispersion and beam depolarization. In principle, one could try to
modify Eq.~(\ref{p_to_f}) to isolate the large-scale contribution
and parameterize the small-scale one by intrinsic Faraday dispersion
and beam depolarization. Wavelets allow us to perform this procedure
in a more straightforward way and to obtain more detailed
information concerning the small-scale component.

\begin{figure}
\begin{picture}(240,225)(0,0)
\put(0,115){\includegraphics[width=0.4\textwidth]{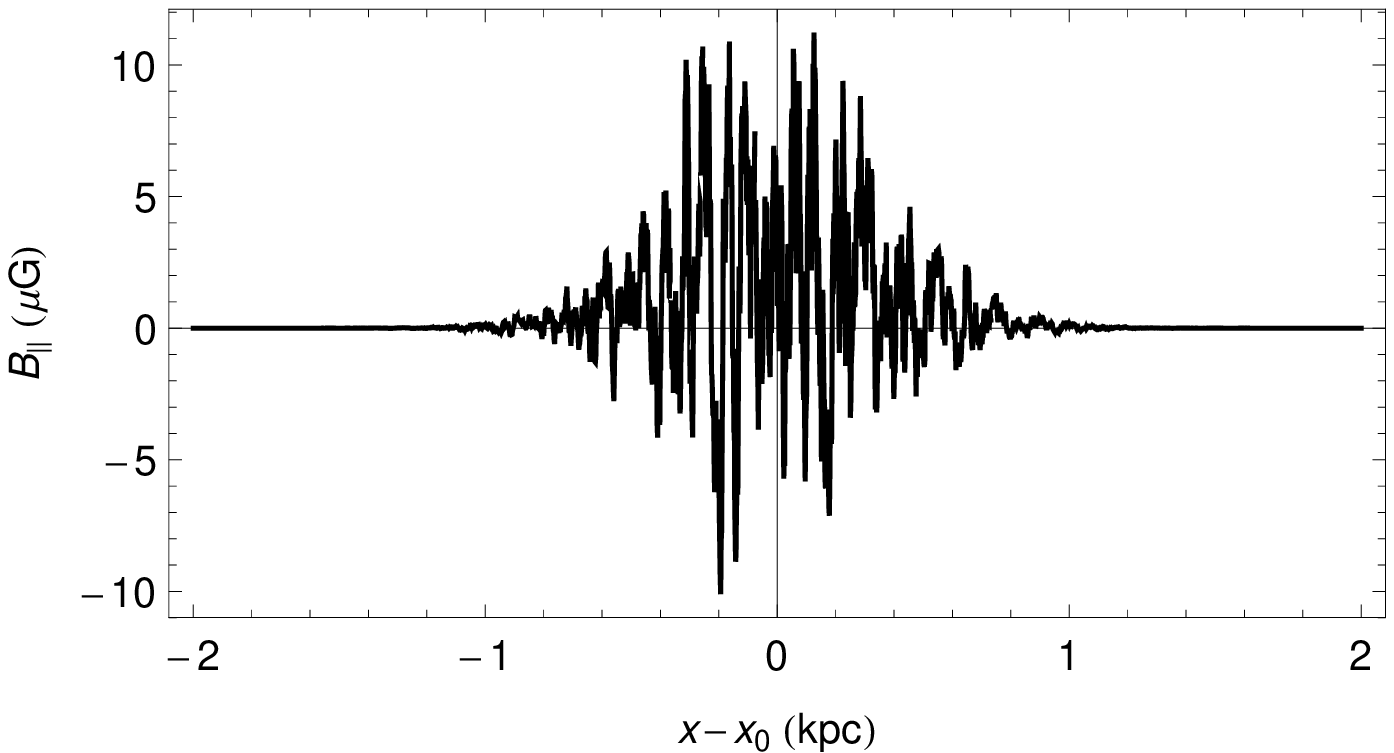}}
\put(0,00){\includegraphics[width=0.4\textwidth]{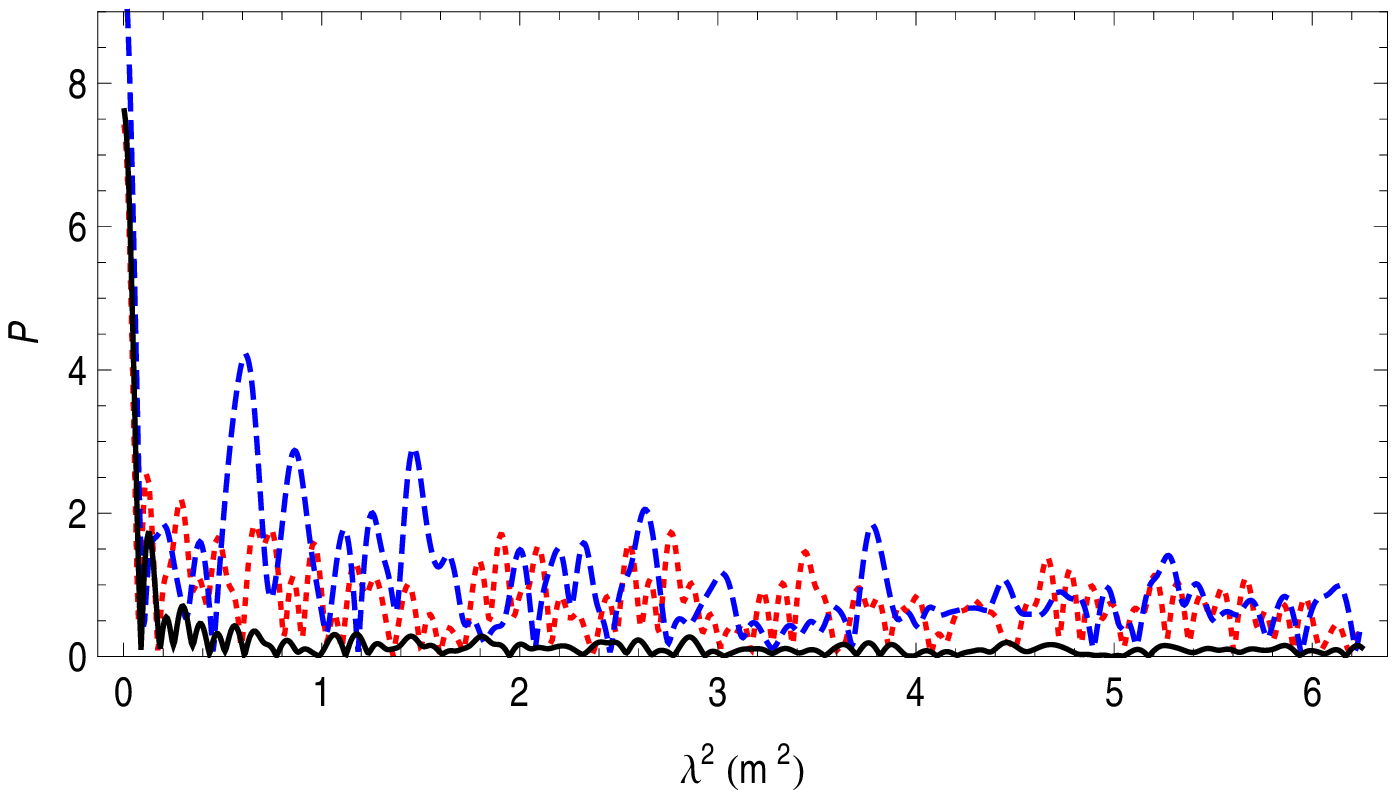}}
\put(210,210){(a)}
\put(210,100){(b)}
\end{picture}
\caption {Magnetic field distributions over a galactic disc. The
ratio of small-scale ($b$) and large-scale ($B$) components is
$b/B=2$, the ratio of spatial scales of the components is $l/2h=
0.1$. Panel (a): light-of-sight magnetic field; panel (b):
polarized intensity. The long and short dashed lines give $P$ for
two different realizations of the turbulent magnetic fields
calculated for a very narrow beam, while the solid line shows $P$
calculated for a beam which contains 50 independent turbulent
cells.} \label{fig7}
\end{figure}

\begin{figure}
\begin{picture}(240,375)(0,0)
\put(0,250){\includegraphics[width=0.40\textwidth]{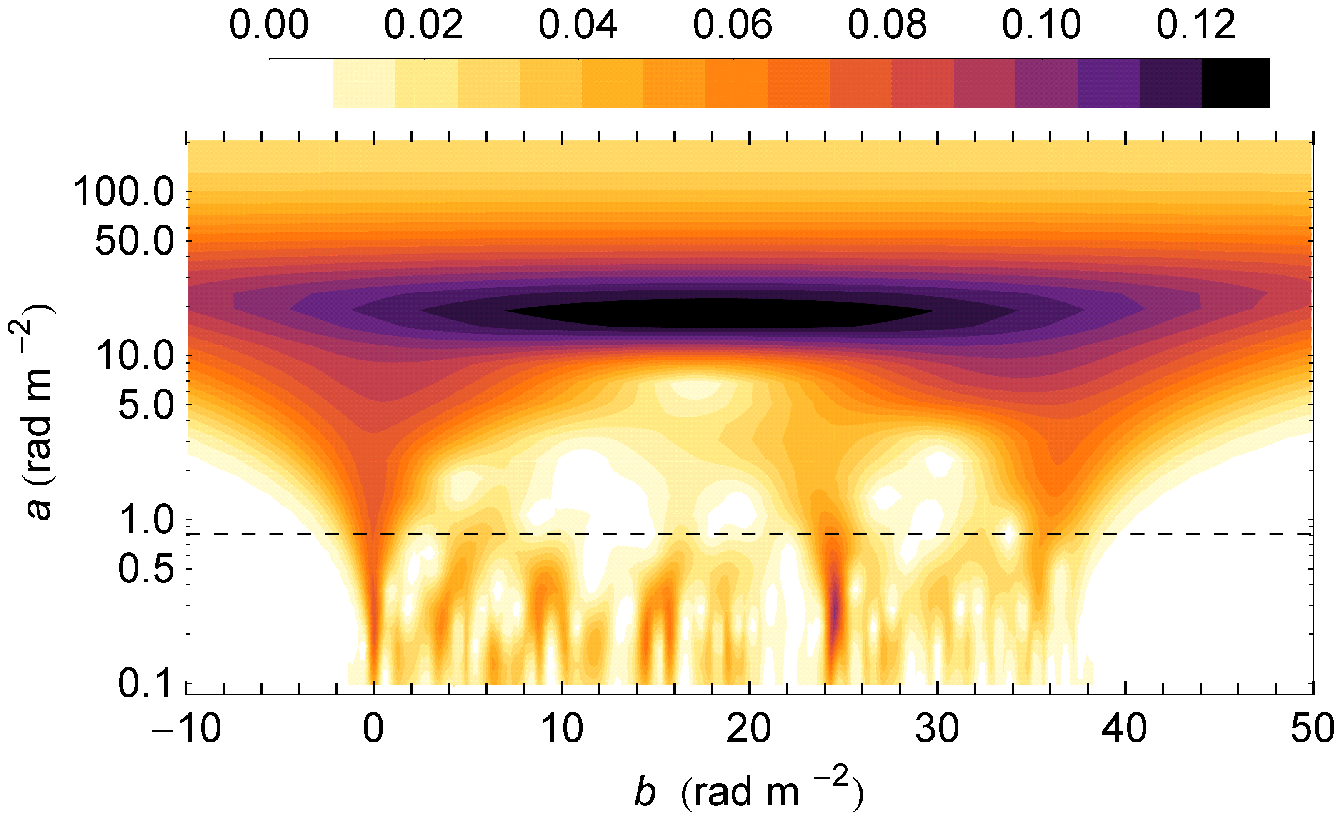}}
\put(0,125){\includegraphics[width=0.40\textwidth]{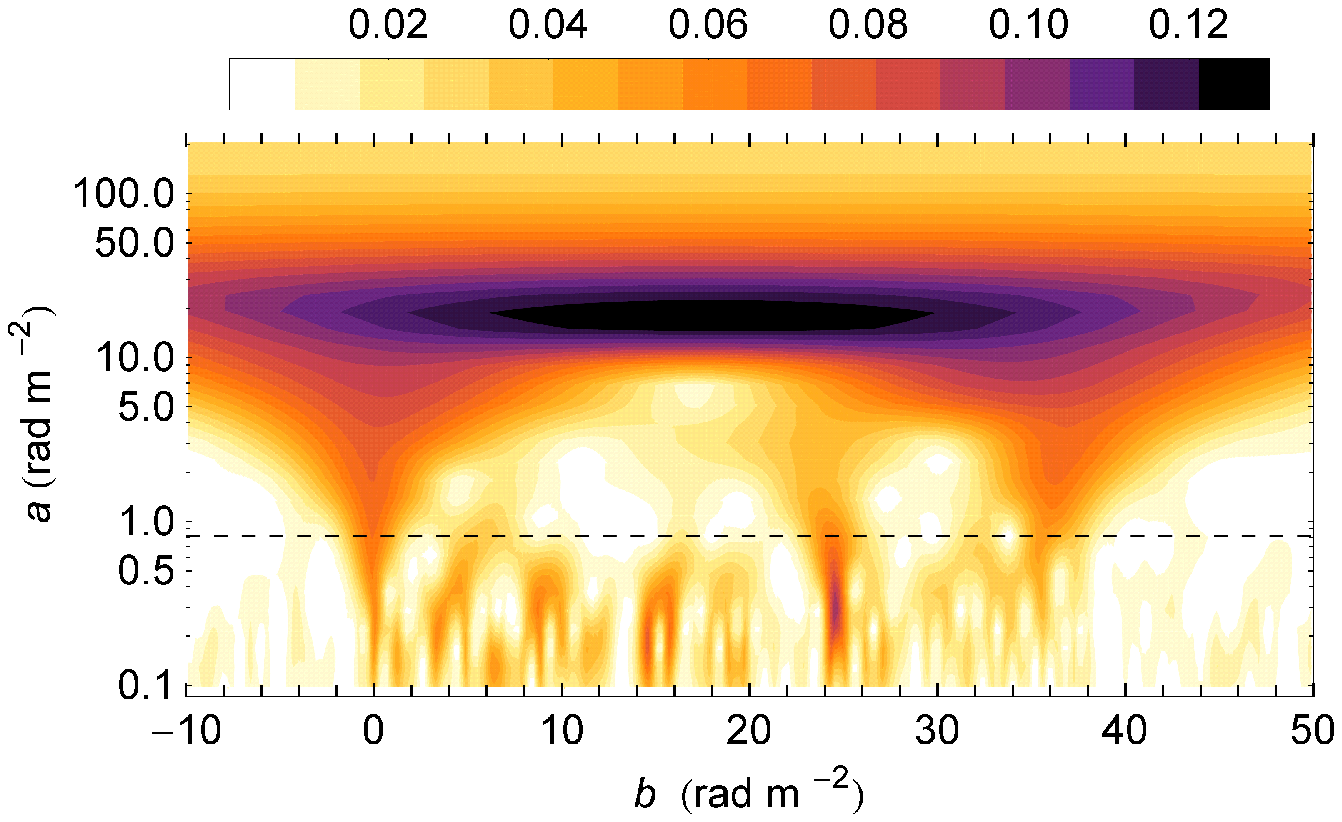}}
\put(0,000){\includegraphics[width=0.40\textwidth]{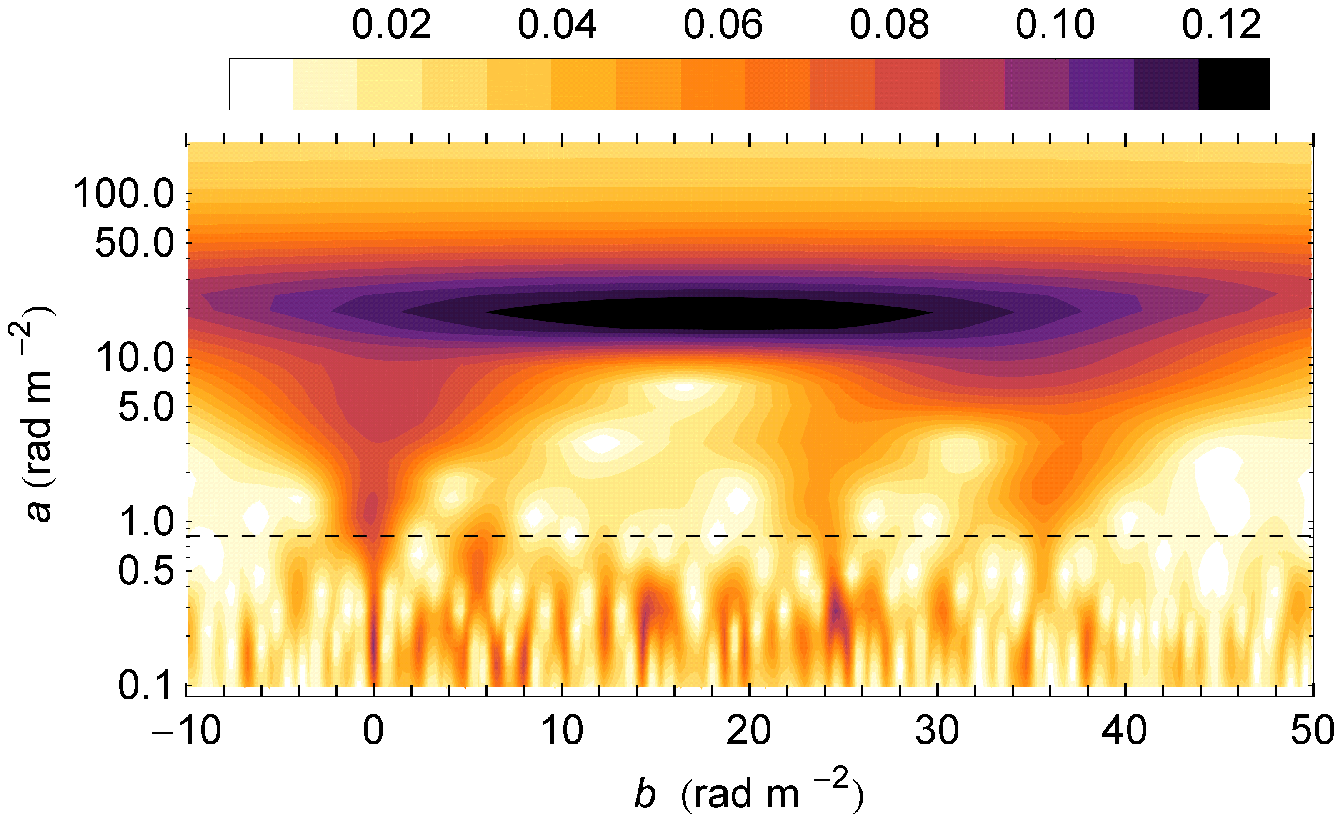}}
\put(210,360){(a)}
\put(210,235){(b)}
\put(210,110){(c)}
\end{picture}
\caption {Wavelet plane $w_F(a,b)$ calculated for different ratios
of signal-to-noise from $P(\lambda^2)$, obtained for the wide beam
(shown in Fig.~\protect{\ref{fig7}}b by the black line). From top to
bottom: signal without noise, signal-to-noise ratios of 2 and
0.5. The horizontal dashed line shows the upper bound of the domain
admissible to LOFAR-type observations. } \label{fig9}
\end{figure}

We start here with an example of a magnetic field which contains
large-scale and turbulent components (Fig.~\ref{fig7}a) and consider
a slab with thickness $2h = 1$\,kpc with a large-scale magnetic
field $B$ and a turbulent magnetic field with a Kolmogorov spectrum
and an rms value $b$ with the scale $l=0.1$\,kpc ($l/2h = 1/20$) and
$b/B = 2$. The polarized intensity obtained from such a slab is
affected by two depolarization effects. One effect known as internal
Faraday dispersion occurs because several (about 20 in our example)
independent turbulent cells are located at a given line of sight
passing through the slab. This effect occurs even for a very narrow
beam, with a cross-section with the slab surface containing just one
turbulent cell. The corresponding $P$ for two independent
realizations of the turbulent field is shown by dashed lines in
Fig.~\ref{fig7}b.

The other depolarization effect arises because the beam is usually
wide enough to contain many turbulent cells. If, say, the beam
diameter on the slab is 1\,kpc it contains about 100 independent
cells. The polarized emission passing through independent cells
contributes to $P$, leading to additional depolarization known as
beam depolarization. The values of $P$ calculated for a beam with a
cross-section containing 50 independent cells is shown by the solid
line in Fig.~\ref{fig7}b. We see that typical $P/P_{\rm max}$ values
are only a few percent for large $\lambda^2$.
 One could conclude
that the emission is almost
completely depolarized and useless for RM Synthesis.  However,
the wavelet method being applied to $P$ shows a well-ordered
structure (Fig.~\ref{fig9}). The point is that wavelets separate
contributions from different spatial scales and associate them with
specific domains in the plot.

In Fig.~\ref{fig9}a, we see a contribution from the large-scale
magnetic field (the long horizontal feature at the upper part of the
panel) and various small-scale details at the bottom which represent
the contributions of individual turbulent cells. Having a limited
spectral range covered by the observations, we can obtain wavelet
coefficients in a limited part of the wavelet plane only. The border
of this part is given by Eq.~(\ref{crit}) which reads as
\begin{equation}
a_{\rm max} = \lambda_{\rm min}^{-2}.
 \label{acrit}
 \end{equation}
Here we take into account that for structures of size $a$ the range
of Faraday depth is of order $\Delta \phi = a^{-1}$. The upper limit
of the domain admissible for LOFAR observations is shown in
Fig.~\ref{fig9} by a horizontal dashed line. We conclude that the
contribution of the large-scale magnetic field remains inaccessible
for this type of observations while the small-scale details remain
visible. Note that the wavelet transform presented in the figure
allows us to recognize the range of Faraday depths in which
small-scale structures occur (from 0\,rad m$^{-2}$ to 38 \,rad
m$^{-2}$ for the given example). Obviously, using data at large
$\lambda^2$ only, we cannot confirm (or reject) the existence
of a large-scale magnetic field.

The method under discussion presumes that the signal-to-noise ratio
is sufficiently large. The lower panels of Fig.~\ref{fig9}b present
the result of the wavelet RM Synthesis for $P$ for which random
instrumental noise (independent in each spectral channel from the
noise in the other channels) is added. The signal-to-noise ratio
(calculated from the averaged signal and the noise per frequency
channel and
per $Q$ and $U$ point for the observational range $1.25 < \lambda <
2.5$\, m) is about 2 (panel b) and the contribution of the
small-scale fields is easily distinguishable from
instrumental noise. These plots shows that a signal-to-noise ratio about
2  (per channel and per $Q$ and $U$ point)
or higher is sufficient
to isolate the range in Faraday space responsible for the
small-scale magnetic fields. For illustrative purposes we give an
example with the signal-to-noise ratio of about 0.5 and demonstrate
that the contribution of small-scale field cannot be distinguished
from instrumental noise (Fig.~\ref{fig9}c).

\section{A synthetic Example}
\label{SE}

To summarize our findings and to conclude what can be learned from
polarized radio observations of galactic magnetic fields using
different RM Synthesis techniques, we now consider a complex test
example. Suppose that we observe a galaxy which hosts both regular
and turbulent magnetic fields (like the example considered in
Sect.~\ref{turb}) and a strong point-like source of polarized radio
emission behind the galaxy on the same line of sight. Independent of
its position in physical space (coordinate $x$) behind the galaxy,
this source appears in Faraday space like attached to the galactic
backside. For sake of definiteness we suppose that the contribution
to $F$ which comes from the galaxy is purely real and that one from
the point source is purely imaginary. The task for the analysis is
to reconstruct the Faraday dispersion model and then to
restore the intensity and the polarization angle of the point
source, to separate the contribution of large-scale and small-scale
galactic magnetic field and recognize the shape of the large-scale
structure.

\begin{figure}
\begin{picture}(240,465)(0,0)
\put(0,335){\includegraphics[width=0.40\textwidth]{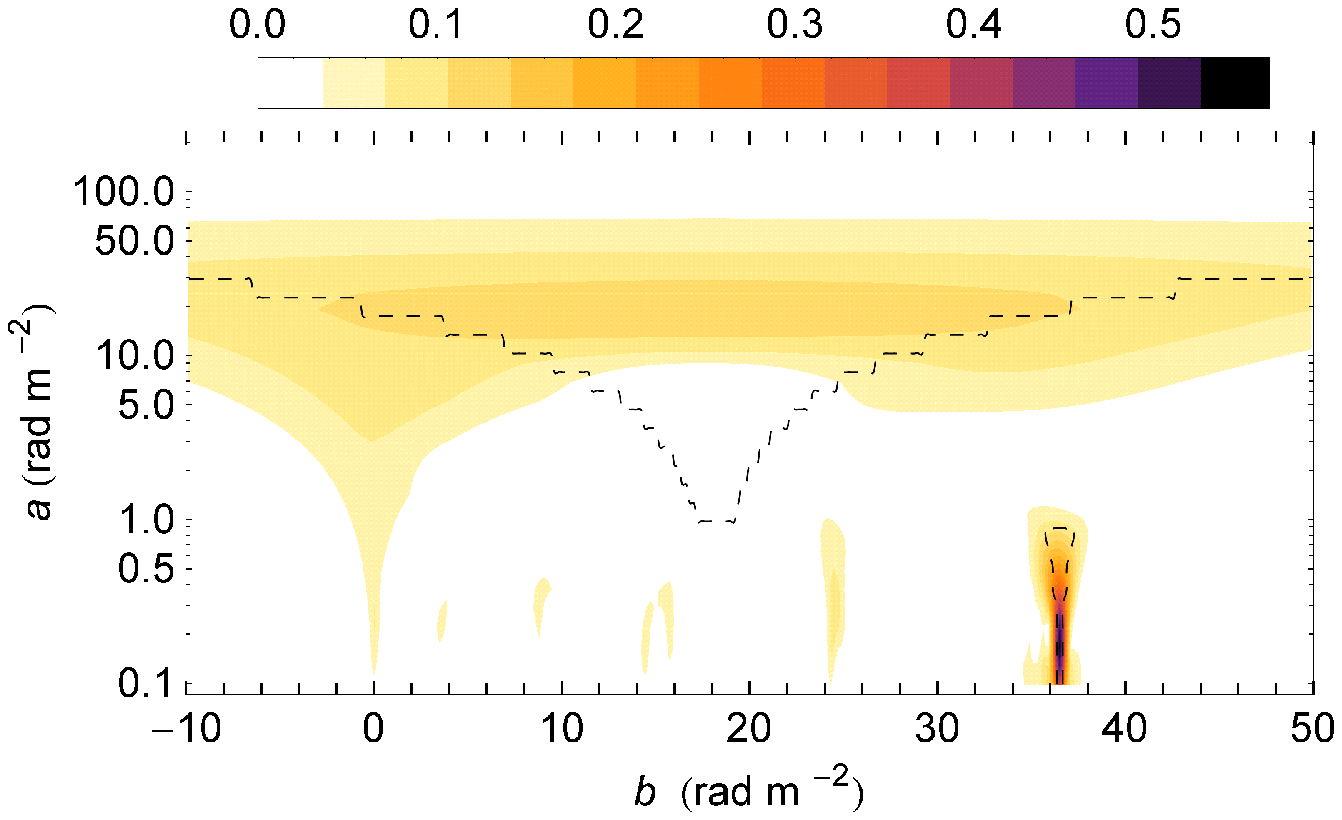}}
\put(0,220){\includegraphics[width=0.40\textwidth]{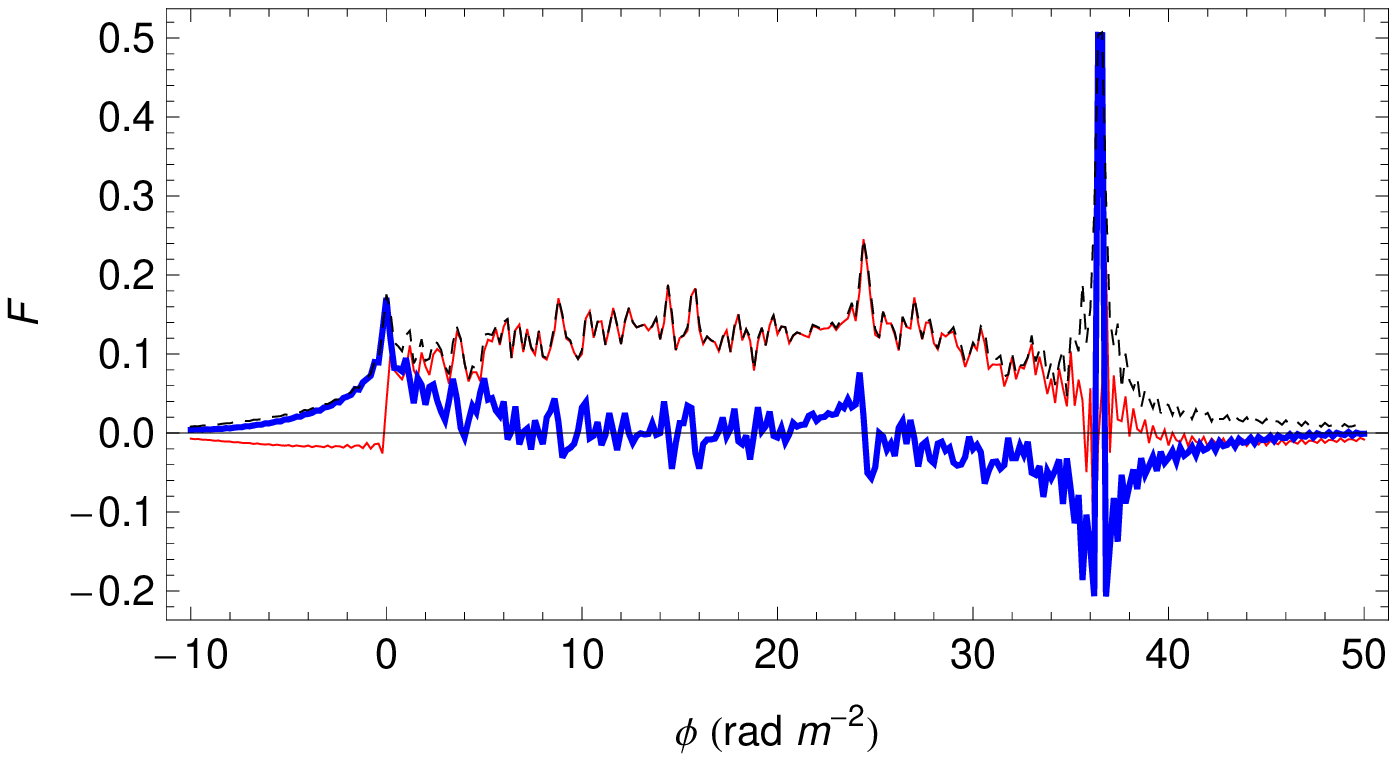}}
\put(0,110){\includegraphics[width=0.40\textwidth]{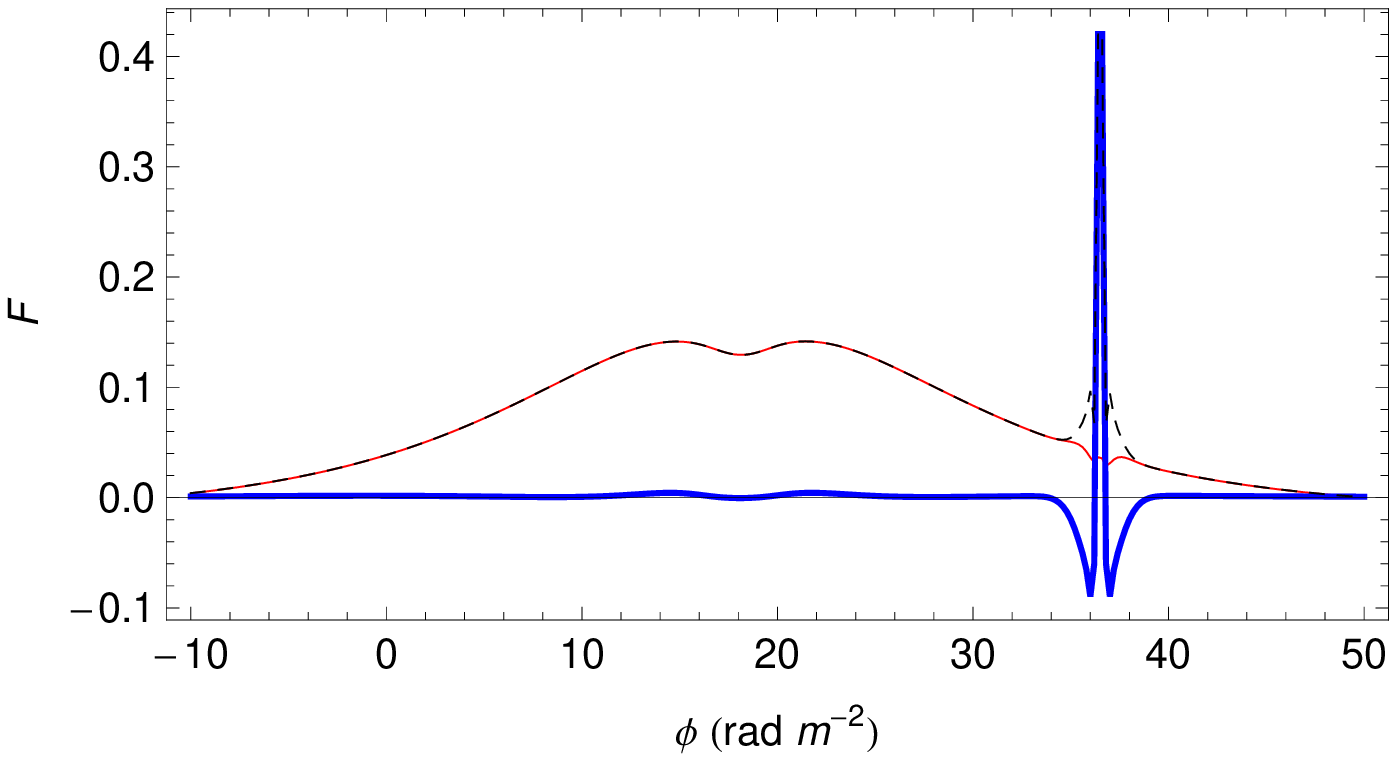}}
\put(0,000){\includegraphics[width=0.40\textwidth]{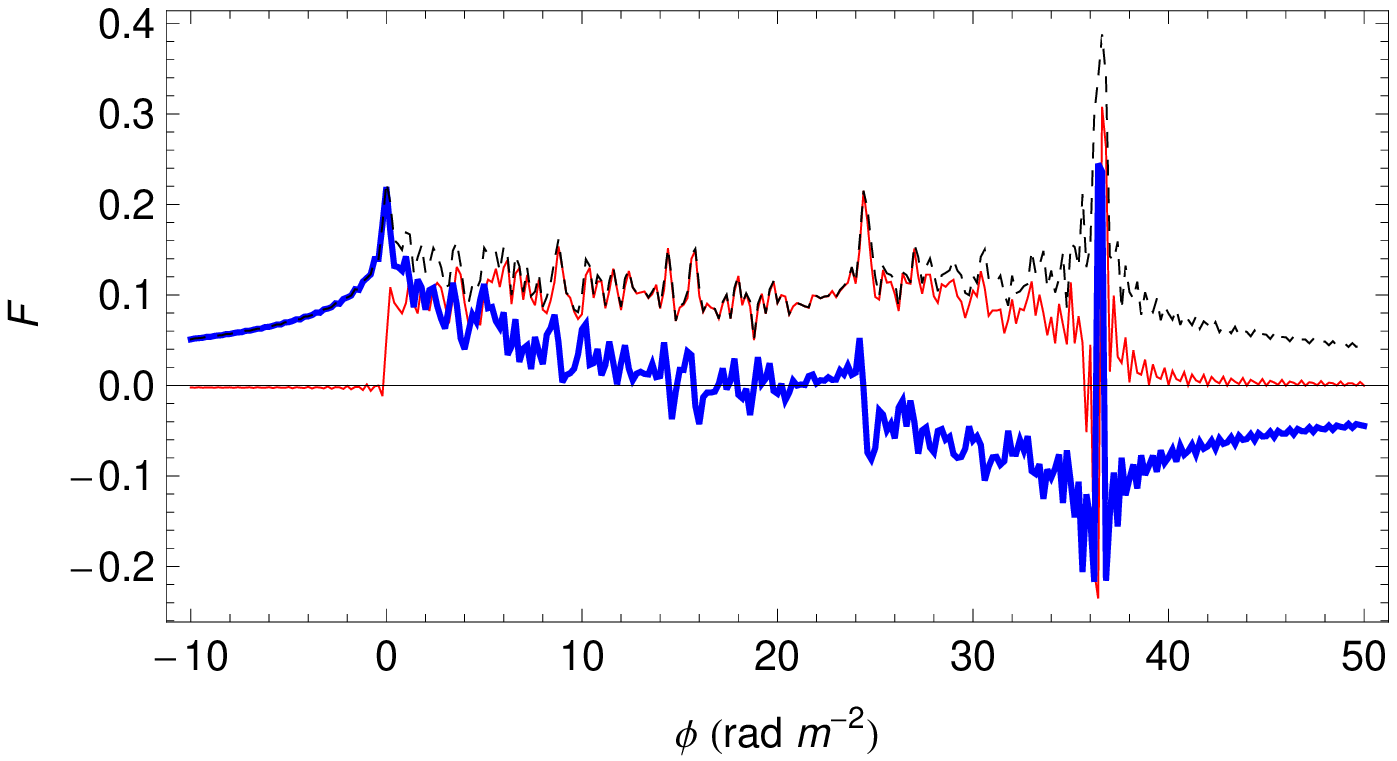}}
\put(210,435){(a)} \put(210,320){(b)} \put(210,210){(c)}
\put(210,100){(d)}
\end{picture}
\caption {Panel (a): Wavelet plane $w_F(a,b)$ (in
colors online) calculated from the polarized radio emission
$P(\lambda^2)$ generated by turbulent galactic magnetic fields
superimposed on a point source behind the galaxy. The results of RM
Synthesis are shown in panels (b)--(d). Panel (b): wavelet
reconstruction using the whole wavelet plane; panel (c): wavelet
reconstruction using only the domains marked in the wavelet plane by
dashed lines in panel (a); panel (d): reconstruction by standard RM
Synthesis. The red thin line is for the real part, the blue thick line for the
imaginary part, and the black dashed line for the modulus. }
\label{fig_last}\end{figure}

Fig.~\ref{fig_last}a shows the wavelet plane calculated for this
signal. We recognize here an extended structure in the region of
large scales, $10\lesssim a \lesssim 50$ rad m$^{-2}$, which we
identify with the contribution of a galactic disc as a whole,
small-scale details at $a < 1$ rad m$^{-2}$, which we identify as
the contributions of the small-scale magnetic fields, and a bright
horn located at $b \approx 37$ rad m$^{-2}$. Then we perform the
wavelet-based RM Synthesis using the symmetry arguments to the
galactic contribution with the reflection point at $\phi_0^g=18$ rad
m$^{-2}$ and considering the domain of the horn as a point-like
source, i.e. apply the symmetry condition at the point $\phi_0^h=37$
rad m$^{-2}$. This means
\begin{eqnarray}
\label{phi0}
 \phi_0(a,b) =
\left\{
  \begin{array}{ll}
   \phi_0^g, & a > 1 \\
   \phi_0^h, & a\leq 1
  \end{array}.
\right.
\end{eqnarray}
Applying the symmetry arguments for
the whole domain would amplify the small-scale noise and reflect it
with respect to the position of the maximum. Therefore we perform
a reflection locally (only in vicinity of the maxima) which leads to the definition
\begin{eqnarray}
\label{w-w+n} w_-(a,b) =
\left\{
  \begin{array}{ll}
   w_{+}\left(a,2\phi_0(a,b)-b \right), & |b-\phi_0(a,b)|\leq a \\
   0, & |b-\phi_0(a,b)|> a
  \end{array}
\right.
\end{eqnarray}
instead of Eq.~(\ref{w-w+}). The dashed lines in
Fig.~\ref{fig_last}a shows two domains around each maxima  in which the symmetry
arguments are used for calculation of $w_{-}$  following Eqs.~(\ref{phi0}--\ref{w-w+n}).

The reconstruction in Fig.~\ref{fig_last}b shows a point-like source
and a smooth contribution from the large-scale magnetic field,
superimposed onto fluctuations associated with the small-scale
fields. If we are not interested in the small-scale details we
ignore the wavelet coefficients outside isolated domains and obtain
a smoothed $F$ which represents the contribution of the mean field
in the disc and the point-like source only (Fig.~\ref{fig_last}c).
We see a smoothed real contribution from the mean field and an
imaginary point-like source, i.e. the method correctly reproduces
the polarization angles. For comparison with
Fig.~\ref{fig_last}b, we show in Fig.~\ref{fig_last}d the result
obtained by the standard RM Synthesis technique where the
phase information is completely lost.  More specifically, the blue thick
line at Fig.~\ref{fig_last}b follows the horizontal axis as required
by the model while  a negative trend of the blue thick line is visible in
Fig.~\ref{fig_last}d.

\section{Discussion and Conclusions}
\label{DC}

We considered how to apply RM Synthesis to extract information
concerning magnetic fields of spiral galaxies from polarized
emission data. The main attention was given to two related novelties
in the method, i.e. the symmetry argument and the wavelet technique.
Symmetry arguments open the possibility to reconstruct
$P(\lambda^2)$ for negative $\lambda^2$ from $P(\lambda^2)$ for
positive $\lambda^2$. If $F$ contains several contributions (say,
the disc and an unresolved source) the reconstruction can be
performed locally in Faraday depth space using wavelets. The wavelet
technique appears to be useful in RM Synthesis in several areas
which are not directly related to the problem of negative
$\lambda^2$. In particular, they allow us to perform RM Synthesis
locally in Faraday depth space.

The idea of RM Synthesis  was applied first  for
sources which are presumed to be
point-like in Faraday space or {a combination of several point-like
sources}  \citep{
2003A&A...403.1031H,2005A&A...441..931D,2009A&A...503..409H, 2010AJ....139.1681W,2010A&A...520A..80L}.  We demonstrated that the traditional RM
Synthesis for a point-like source indirectly  implies a symmetry
argument  (see Eq. 22) and, in this sense, can be considered as
a particular case of the method under discussion here.

Investigating the applications of RM Synthesis to the polarization
details associated with small-scale magnetic fields, we isolated an
option which was not covered by \cite{Burn1966MNRAS.133...67B}.
\cite{Burn1966MNRAS.133...67B} (see also \cite{1998MNRAS.299..189S})
described a contribution of small-scale fields in terms of Faraday
dispersion and beam depolarization, i.e. using averages over a
corresponding statistical ensemble. Here
we  use the complex polarization $P$ for RM Synthesis without any
averaging and demonstrate that it allows us to obtain ( if the
$\lambda^2$ coverage is adequate) a range in Faraday
space where the contribution from small-scale field is located.

A general conclusion concerning the applicability of RM Synthesis to
the interpretation of the radio polarization data for extended
sources like spiral galaxies is that quite severe requirements are
needed to provide the full applicability of the method. The
most severe one is the requirement for the wavelength range where
short wavelengths are highly desired. \cite{2009IAUS..259..669B}
show that RM Synthesis applied at the border of its applicability
(say, if the number of frequency channels is very small) can
sometimes be misleading.

Traditional methods of the pattern recognition for magnetic
structures in spiral galaxies (e.g.
\cite{1990A&A...230..284R,1992A&A...264..396S,2000MNRAS.318..925F,
2004A&A...414...53F,2006A&A...458..441P,2008A&A...480...45S}) were
mainly based on the interpretation of polarization angles at a few
wavelengths only. The
efforts of the above papers concentrated mainly on the
distribution of polarization angles in the radio image. Modern
wide-band observations with many frequency channels and RM Synthesis
open wide perspectives to extract much more information from the
polarization signal in a given
telescope beam. However, they cannot
fully replace the traditional analysis of the distribution of the
polarization patterns. Of course, the development of RM Synthesis in
order to deal with radio images as a whole rather than with just
one beam
is a goal for further developments, and wavelets are
likely to be helpful here. However, this is outside of the scope of
this paper.

Obviously, a quantified characteristic of reconstruction fidelity of the algorithm under discussion is highly desirable. However, a general norm is not informative in a spectral problems with varying windows -
 one can completely lose the large-scale structures and reconstruct precisely the small-scale details (or vise versa). Fig.~\ref{fig66}  is a good illustration for that: the central part of the structure is completely lost (for $0.06<\lambda<0.42$), but the borders are reconstructed quite well. Thus the "global fidelity" is very poor, while the "fidelity of border reconstruction" is fine.
 So, we avoided intentionally to quantify the
quality of the reconstruction by a norm in some functional
space.

The RM Synthesis approach was suggested in the context of
observations with a rather large value of $\lambda_{\rm min}$ while
the traditional method to measure RMs between only two wide
frequency bands intensively  used the data at $\lambda=3$\,cm
and $\lambda=6$\,cm and rarely used the wavelength range $\lambda >
22$\,cm. The attention of the traditional method was focused on the
magnetic field configuration rather than on the Faraday dispersion
function which appears in this context as an intermediate rather
than a final result. On the other hand, observations in the
wavelength range $1.25 < \lambda < 2.5$\,m (the LOFAR high band)
cannot recognize a lot of structures which would be accessible to
observations in the wavelength range $0.03 < \lambda < 2.5$\,m with
an adequate coverage. However, this option
will have to wait for future radio telescopes like the SKA.

This paper, although based on model examples, presents an
optimistic view that the combination the ideas of traditional RM
studies of galactic magnetic fields at short wavelengths and those
of RM Synthesis at large wavelengths is possible. The final conclusion
can be made after application to real data.

\label{lastpage}

\section*{Acknowledgements} This work was supported by the
DFG--RFBR grant 436RUS113/969 (08-02-92881) and by the Council of the President of the Russian Federation (grant MD-4471.2011.1). We thank Dr.
George Heald for providing observational data and for useful
discussions. Our special thanks go to our referee, Dr. Michiel
Brentjens, for his efforts to improve the paper.

\bibliographystyle{mn2e} 
\bibliography{ref}

\end{document}